\documentclass[superscriptaddress,amsmath,amssymb,onecolumn,amsfonts]{revtex4-2}
\usepackage{graphicx}
\graphicspath{{Pictures/}}
\usepackage{hyperref}
\hypersetup{pdfnewwindow=true, colorlinks=true, linkcolor=blue, anchorcolor=blue, citecolor=red, filecolor=blue, menucolor=blue, urlcolor=magenta}
\usepackage{cleveref}
\usepackage{color}
\usepackage{float}
\usepackage{soul}
\usepackage{ulem}
\usepackage{braket}
\usepackage{enumitem}
\usepackage{marginnote}

\newcommand{\sn}{{\rm sn}}
\newcommand{\cn}{{\rm cn}}
\newcommand{\dn}{{\rm dn}}

\newcommand{\sech}{{\rm sech}}

\begin{document}

\title{Superposed periodic kink and pulse solutions of coupled nonlinear equations}

\author{Avinash Khare}
\email{avinashkhare45@gmail.com}
\affiliation{Physics Department, Savitribai Phule Pune University, Pune 411007, India}
\author{Saikat Banerjee}
\email{saikatb@lanl.gov}
\affiliation{Theoretical Division, T-4, Los Alamos National Laboratory, Los Alamos, New Mexico 87545, USA}
\author{Avadh Saxena}
\email{avadh@lanl.gov}
\affiliation{Theoretical Division and Center for Nonlinear Studies, Los Alamos National Laboratory, Los Alamos, New Mexico 87545, USA}
\date{\today}

\begin{abstract}
We present novel previously unexplored periodic solutions, expressed in terms 
of Jacobi elliptic functions, for both a coupled $\phi^4$ model and a coupled 
nonlinear Schr\"odinger equation (NLS) model. Remarkably, these solutions can 
be elegantly reformulated as a linear combination of periodic kinks and 
antikinks, or as a combination of two periodic kinks or two periodic pulse solutions. 
However, we also find that for $m=0$ and a specific value of the periodicity (or at
a nonzero value of the elliptic modulus $m$) this superposition does not hold. 
These results demonstrate that the notion of superposed solutions extends to 
the coupled nonlinear equations as well.
\end{abstract}

\maketitle

\section{Introduction \label{sec:sec.1}}

A striking characteristic of linear theories is the presence of the superposition principle. In linear $n$th order differential equations, there exist $n$ independent solutions, and any other solution can be expressed as a linear combination of these independent solutions. 
However, the highly nontrivial attribute of nonlinear theories arises, in contrast to the linear theories, from the fact that they do not obey the superposition principle, and it is not apparant how many independent solutions exist for a nonlinear equation. Nevertheless, several nonlinear equations, such as $\phi^4$, NLS, Korteweg-de Vries (KdV), and modified KdV (mKdV) etc., have found extensive applications in various areas of physics, ranging from the dynamics of coupled Bose-Einstein condensates~\cite{ac21} to nonlinear photonic integrations~\cite{zw14}. Interestingly, these equations have been shown to exhibit a form of superposition principle~\cite{ks13}. Particularly, it has been demonstrated that if a nonlinear differential equation admits a periodic pulse solution expressed as Jacobi elliptic functions~\cite{as} $\dn(x,m)$ and $\cn(x,m)$, then a superposed periodic solution $\dn(x,m) \pm \sqrt{m} \
cn(x,m)$ will also be allowed with $m$ being the modulus of the Jacobi elliptic function. In a similar fashion, if a nonlinear equation admits a solution expressed as $\dn^2(x,m)$, then another solution expressed as $\dn^2(x,m) \pm \sqrt{m} \dn(x,m) \cn(x,m)$ will also be allowed. Notably, most of the nonlinear differential equations mentioned above also exhibit a different kind of superposition principle. Specifically, it has been demonstrated that if a nonlinear differential equation admits $\dn(x,m)$ and/or $\cn(x,m)$ as solutions, then it will also allow for complex parity-time (PT)-invariant solutions expressed as $\dn(x,m) \pm i \sqrt{m} \sn(x,m)$ and/or $\cn(x,m) \pm i \sn(x,m)$~\cite{ks16,ks18,ks19}. In a similar way, one finds that if there is a periodic kink solution in terms of $\sn(x,m)$ for a given nonlinear differential equation, then it also allows for a complex PT-invariant solution expressed as $\sqrt{m} \sn(x,m) \pm i \dn(x,m)$ in addition to the form $\sn(x,m) \pm i \cn(x,m)$. Moreover, if a nonlinear equation hosts a solution expressed as $\dn^2(x,m)$, it will also allow for complex PT-invariant periodic solutions, represented as $\dn^2(x,m) \pm i \sqrt{m} \dn(x,m) \sn(x,m)$ or $\dn^2(x,m) \pm i m\sn(x,m) \cn(x,m)$~\cite{ks16,ks18,ks19}.
 
Inspired by a previous theoretical work~\cite{per}, Tankeyev, Smagin, Borich, and Zhuravlev~\cite{tan,sma} obtained a novel solution that can be expressed as a combination of a kink and an antikink solution. Analogous solutions have been previously obtained in the realms of condensed matter physics and field theory~\cite{dashen,campbell,saxena,thies}. The key step in the work of Tankeyev et al. \cite{tan} lies in their utilization of a hyperbolic identity. Building upon their findings, we decided to explore similar identities related to the Jacobi elliptic functions $\sn(x,m)$, $\cn(x,m)$, and $\dn(x,m)$, which guided us in discerning the potential structure of superposed periodic solutions. In our earlier investigation~\cite{ks22a}, we showed that a wide class of nonlinear differential equations, such as the the symmetric and asymmetric $\phi^4$ equation, the NLS, the quadratic-cubic NLS, mKdV and mKdV-KdV equations admit superposed periodic kink and pulse solutions, with some cases also admitting superposed hyperbolic kink solutions. Subsequently, in another notable study~\cite{ks22b}, we presented novel superposed hyperbolic solutions for few coupled nonlinear equations, in terms of kink-antikink, or two kinks or two pulse solutions.

Hence, a natural question arises: Can the aforementioned coupled equations also admit periodic solutions which can be reexpressed as superposition of periodic kink-antikink or two periodic kinks or two periodic pulse solutions? In this paper we show that this is indeed the case. Specifically, we extend our investigation to the same coupled $\phi^4$ model and the coupled NLS model explored in~\cite{ks22b} and show that both these models admit periodic solutions, which can be reexpressed as either the superposition of a periodic kink and an antikink, or two distinct periodic kinks, or two periodic pulse solutions.

Throughout this paper a periodic solution refers to a solution that exhibits spatial periodicity, unless stated otherwise. Furthermore, we employ the term ``periodic kink" (or ``periodic pulse") solution to denote a kink (or pulse) lattice solution that goes over to the hyperbolic kink (or pulse) solution in the limit of $m = 1$. Finally, we note that so far we have been unable to obtain periodic solutions of the coupled mKdV equation, which can be expressed as a superposition of a periodic kink and an antikink, or two periodic kinks, or two periodic pulse solutions.

The paper is structured as follows. In Section~\ref{sec:sec.2}, we focus on the same coupled $\phi^4$ model examined in~\cite{ks22b} and obtain a large number of new solutions that can be reexpressed as a superposition of either a periodic kink and an antikink, or two periodic kinks, or two periodic pulse solutions. In Subsection~\ref{sec:sec.3a} and \ref{sec:sec.3b} respectively, we explore the same solutions but in the special cases of $B = \frac{m}{1-m} > 0$ as well as $m=0$. In these cases, the solutions cannot be reexpressed as a superposition of two periodic kinks, or a periodic kink-antikink pair, or two periodic pulse solutions. Notably, we find that some of these solutions in case $B = \frac{m}{1-m}$ exhibit nonreciprocal behavior. Moving on to Section~\ref{sec:sec.4}, we delve into the coupled nonlinear Schrödinger equation (NLS) model discussed in~\cite{ks22b}. Interestingly, we establish a mapping between this model and the coupled $\phi^4$ model discussed in Section~\ref{sec:sec.2}, allowing us to readily extract the superposed solutions for the NLS model. Lastly, in Section~\ref{sec:sec.6}, we summarize our main findings and discuss a few open problems. In Appendix~\ref{sec:app.1}, we discuss the superposed solutions of the coupled $\phi^4$ model under the assumption that the two fields are proportional to each other and show that these solutions can be reexpressed as a superposition of either a periodic kink and an antikink, two periodic kinks, or two periodic pulse solutions (specifically, of $\dn(x,m)$ type). In Appendix~\ref{sec:app.2} we obtain six solutions involving Lamé polynomials of order two, three solutions with other Lam\'e polynomials of order two, and three nonreciprocal solutions involving Lam\'e polynomials of order one. 


\section{A Coupled $\phi^4$ Model \label{sec:sec.2}}

Let us consider the following coupled $\phi^4$ model characterized by the equations~\cite{ks22b} 
\begin{subequations}
\begin{align}
\label{eq.1.1}
\phi_{1xx} & = a_1 \phi_1 +(b_1 \phi_{1}^{2}+ d_1 \phi_{2}^2)\phi_1, \\
\label{eq.1.2}
\phi_{2xx} & = a_2 \phi_2 +(b_2 \phi_{1}^{2}+ d_2 \phi_{2}^2)\phi_2 .
\end{align}
\end{subequations}
Note that these coupled equations can be derived from an interaction potential $V(\phi_1,\phi_2)$ only if $d_1 = b_2$. Before we discuss the coupled superposed periodic kink and pulse solutions, let us note that these coupled equations also admit several periodic kink and pulse solutions most of which are valid even when $b_2 = d_1$.

Next, we show that there are 19 periodic solutions (fifteen in this section and four in Appendix~\ref{sec:app.1}) for these coupled equations. All of them can be expressed as distinct superposed solutions either of the form $\cn(\beta x+\Delta) \pm \cn(\beta x -\Delta)$\,, ~or 
$\dn(\beta x+\Delta) \pm \dn(\beta x -\Delta)$\,, ~or $\sn(\beta x+\Delta) \pm \sn(\beta x -\Delta)$\,. To this end, we will make use of certain identities satisfied by these superposed combinations. These identities are readily derived by
using the following addition theorems for $\sn(x,m)$, $\cn(x,m)$, and $\dn(x,m)$~\cite{as}
\begin{subequations}
\begin{align}
\label{eq.2.1}
\sn(a+b,m) & = \frac{\sn(a,m)\cn(b,m)\dn(b,m)+\sn(b,m)\cn(a,m)\dn(a,m)} {1-m\sn^2(a,m)\sn^2(b,m)} , \\
\label{eq.2.2}
\cn(a+b,m) & = \frac{\cn(a,m)\cn(b,m)-\sn(a,m)\dn(a,m)\sn(b,m)\dn(b,m)} {1-m\sn^2(a,m)\sn^2(b,m)} , \\
\label{eq.2.3}
\dn(a+b,m) & = \frac{\dn(a,m)\dn(b,m)-m \sn(a,m)\cn(a,m)\sn(b,m)\cn(b,m)} 
{1-m\sn^2(a,m)\sn^2(b,m)} ,
\end{align}
\end{subequations}
where $m$ is the modulus of the Jacobi elliptic functions~\cite{as}. In particular, on using Eq.~\eqref{eq.2.1} we obtain the identities
\begin{subequations}
\begin{align}
\label{eq.3.1}
\sn(y+\Delta,m)+\sn(y-\Delta,m) & = \frac{2\sn(y,m) \cn(\Delta,m)} {\dn(\Delta,m)[1+B\cn^2(y,m)]}, \\
\label{eq.3.2}
\sn(y+\Delta,m)-\sn(y-\Delta,m) & = \frac{2\cn(y,m) \dn(y,m)\sn(\Delta,m)} {\dn^2(\Delta,m)[1+B\cn^2(y,m)]},
\end{align}
\end{subequations}
where
\begin{equation}\label{eq.4}
B = \frac{m\sn^2(\Delta,m)}{\dn^2(\Delta,m)} > 0.
\end{equation}
On the other hand, on using Eq.~\eqref{eq.2.2} we obtain the identities
\begin{subequations}
\begin{align}
\label{eq.5.1}
\cn(y+\Delta,m)+\cn(y-\Delta,m) & = \frac{2\cn(y,m) \cn(\Delta,m)} {\dn^2(\Delta,m)[1+B\cn^2(y,m)]}, \\
\label{eq.5.2}
\cn(y-\Delta,m)-\cn(y+\Delta,m) & = \frac{2\sn(y,m) \dn(y,m)\sn(\Delta,m)} {\dn(\Delta,m)[1+B\cn^2(y,m)]}, 
\end{align}
\end{subequations}
where $B$ is again given by Eq.~\eqref{eq.4}. Finally, on using Eq.~\eqref{eq.2.3} we obtain the identities
\begin{subequations}
\begin{align}
\label{eq.6.1}
\dn(y+\Delta,m)+\dn(y-\Delta,m) & = \frac{2\dn(y,m)} {\dn(\Delta,m)[1+B\cn^2(y,m)]}, \\
\label{eq.6.2}
\dn(y-\Delta,m)-\dn(y+\Delta,m) & = \frac{2m \sn(y,m) \cn(y,m) \sn(\Delta,m) \cn(\Delta,m)}{\dn^2(\Delta,m)[1+B\cn^2(y,m)]},
\end{align}
\end{subequations}
where $B$ is again as given by Eq.~\eqref{eq.4}.  

There are two different forms of solutions to the coupled equations Eq.~\eqref{eq.1.1} and Eq.~\eqref{eq.1.2} depending on if $\phi_2(x) \propto \phi_1(x)$ or otherwise. Here we elaborate on those solutions where $\phi_2(x)$ and $\phi_1(x)$ are distinct (and not proportional to each other) while those solutions where $\phi_2(x) \propto \phi_1(x)$ are discussed in Appendix~\ref{sec:app.1}. 

\subsection{Solutions where $\phi_2(x)$ and $\phi_1(x)$ are distinct \label{sec:sec.3}}

We now demonstrate that in case $\phi_1$ and $\phi_2$ are distinct (i.e. not proportional to each other) then the coupled equations Eq.~\eqref{eq.1.1} and Eq.~\eqref{eq.1.2} admit not only superposed periodic kink, i.e. $\sn(x,m)$, and periodic $\dn(x,m)$ pulse solutions but also superposed periodic $\cn(x,m)$ pulse solutions except when either $m = 0$ or $B = \frac{m}{(1-m)}$. In both these cases even though there are exact solutions but these are not superposed solutions. In this section we therefore restrict ourselves to $0 < m \le 1$ and further assume that $B \ne \frac{m}{(1-m)}$. In the next section we will discuss the solutions when $m = 0$ as well as when $B = \frac{m}{(1-m)}$;   
we obtain 15 such superposed solutions. As we show in Appendix~\ref{sec:app.1}, in case the two fields $\phi_1$ and $\phi_2$ are proportional to each other, in that case one is only able to obtain superposed periodic kink and pulse solutions of type $\sn(x,m)$ and $\dn(x,m)$, respectively (and not the pulse solutions of the $\cn(x,m)$ type). \\  

\textbf{Solution I}

\begin{figure}[t]
\centering
\includegraphics[width=1\linewidth]{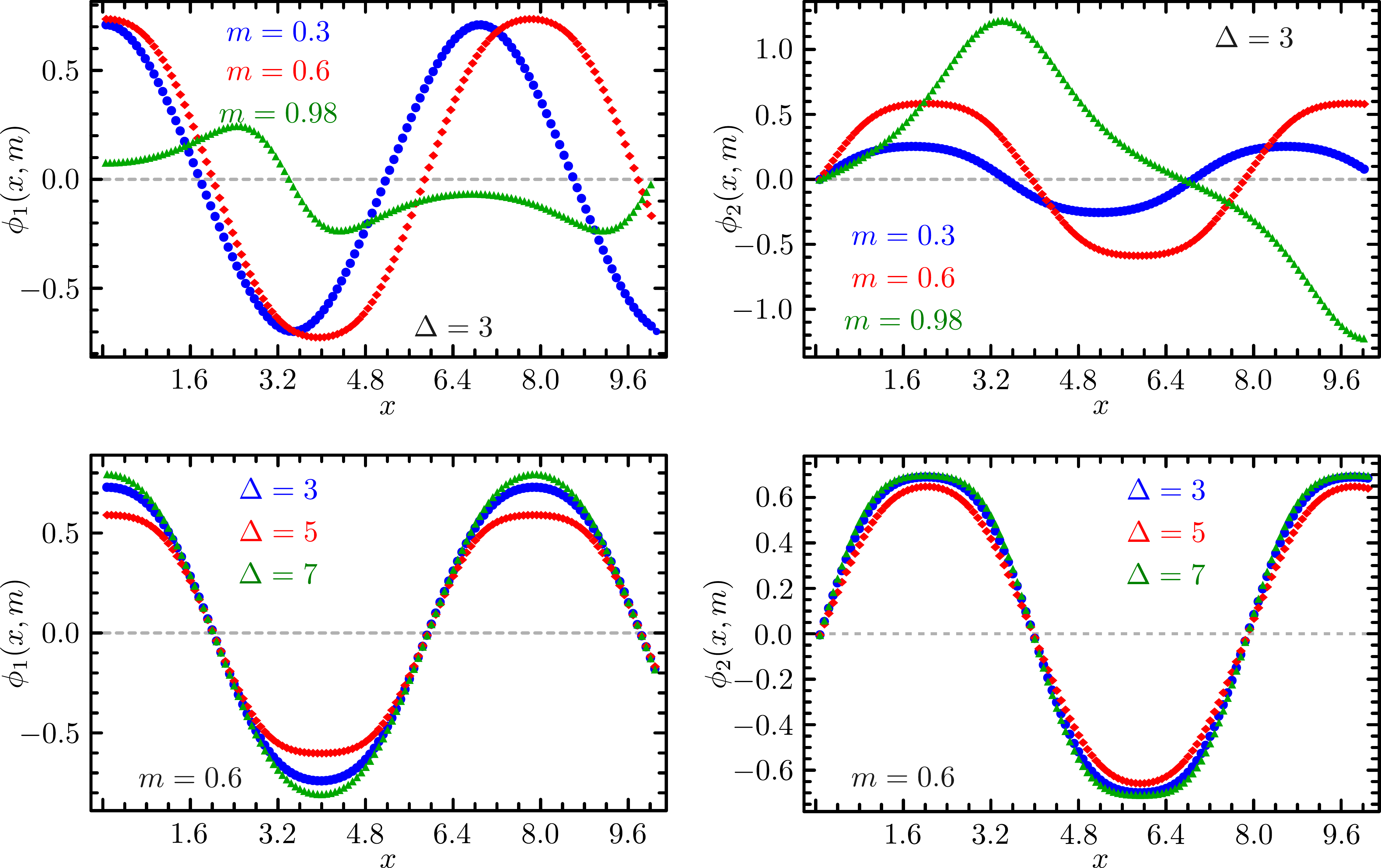} 
\caption{Variations of coupled fields $\phi_1(x,m)$ and $\phi_2(x,m)$ as a function of the position following Eq.~\eqref{eq.7} and Eq.~\eqref{eq.8}, respectively. Note that the scaling parameter $\beta$ is assumed to be 1, and $b_1 = d_2 = -1$.}\label{fig:Fig1}
\end{figure}
It is not difficult to check that
\begin{equation}\label{eq.7}
\phi_1(x) = \frac{A\cn(\beta x,m)}{1+B\cn^2(\beta x,m)}, \quad
\phi_2(x) = \frac{D \sn(\beta x,m)\dn(\beta x,m)}{1+B\cn^2(\beta x,m)}, 
\quad
B > 0,
\end{equation}
is an exact solution of the coupled equations Eq.~\eqref{eq.1.1}, and Eq.~\eqref{eq.1.2}, provided
\begin{align}
\nonumber
&	a_1 = a_2 = (2m-1)\beta^2, \quad d_1 D^2 = 3 d_2 D^2 = -6 B\beta^2, \\
\label{eq.8}
&	b_2 A^2  = 3 b_1 A^2 = -6(B+1)[m-(1-m)B]\beta^2 .	
\end{align}
Since $B > 0$, it then follows that for this solution $d_1, d_2 < 0$. On the other hand, $b_1, b_2 > (< )$ 0 depending on if $ m < (> )$ $(1-m)B$. Finally $a_1, a_2 \ge (<)$ 0 depending on if $m \ge (<)$ 1/2. Note that in this section we are considering the solutions with $m \ne (1-m)B$ and $m \ne 0$. Further, in case $b_2 = d_1$, then from Eq.~\eqref{eq.8} it follows that $d_2 = b_1$. We might add that in all the solutions given below, $B > 0$. Further, in all the solutions given below only $A^2$ and $D^2$ appear and without any loss of generality we choose $A, D > 0$.

On using the identities Eq.~\eqref{eq.5.1} and Eq.~\eqref{eq.5.2}, the coupled solution Eq.~\eqref{eq.7} can be re-written as 
\begin{subequations}
\begin{align}
\label{eq.9.1}
\phi_1(x) & = \sqrt{\frac{m}{2|b_1|}}\beta [\cn(\beta x+\Delta,m)+\cn(\beta x-\Delta,m)], \\
\label{eq.9.2}
\phi_2(x) & = \sqrt{\frac{m}{2|d_2|}} \beta [\cn(\beta x-\Delta,m)-\cn(\beta x+\Delta,m)],
\end{align}
\end{subequations}
where $B = \frac{m\sn^2(\Delta,m)}{\dn^2(\Delta,m)}$. The structure of these solutions is shown in Fig.~\ref{fig:Fig1} for distinct choices of the shift parameter $\Delta$, and the modulus $m$. We have independently checked this solution and all subsequent solutions using \texttt{MATHEMATICA V.13.1} as well. 

{\bf Solution II}

\begin{figure}[b]
\centering
\includegraphics[width=1\linewidth]{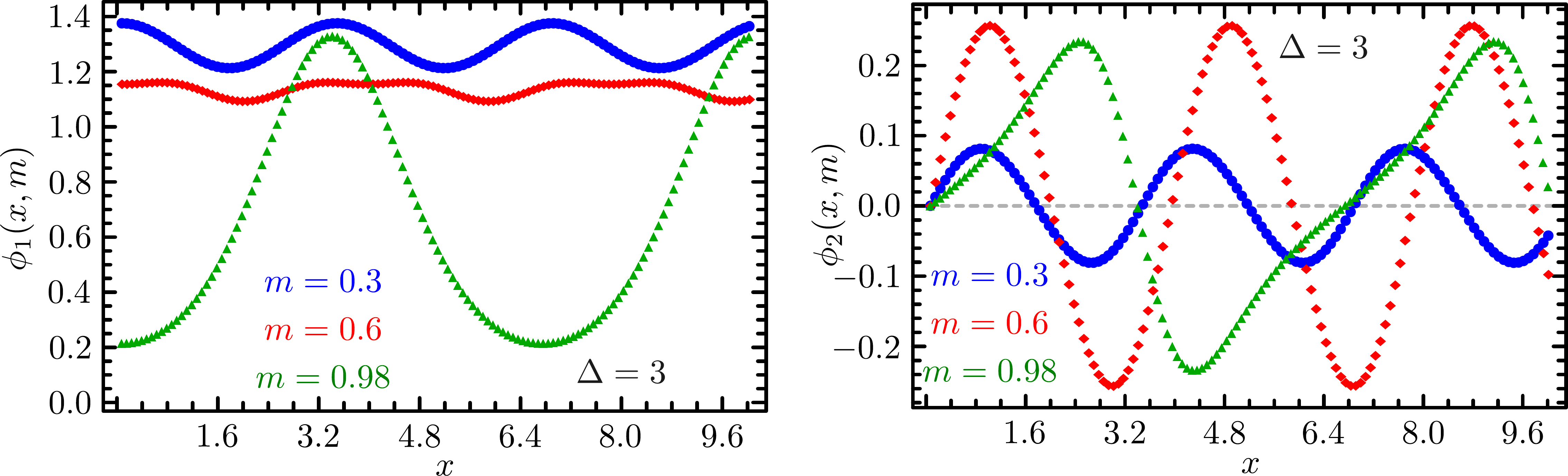} 
\caption{Variations of coupled fields $\phi_1(x,m)$ and $\phi_2(x,m)$ as a function of the position following Eq.~\eqref{eq.10} and Eq.~\eqref{eq.11}, respectively. Note that the scaling parameter $\beta$ is assumed to be 1, and $b_1 = d_2 = -1$.}\label{fig:Fig2}
\end{figure}

We note that
\begin{equation}\label{eq.10}
\phi_1(x) = \frac{A\dn(\beta x,m)}{1+B\cn^2(\beta x,m)}, \quad
\phi_2(x) = \frac{D \sn(\beta x,m) \cn(\beta x,m)} {1+B\cn^2(\beta x,m)}, \quad
 \end{equation}
is an exact solution of the coupled equations Eq.~\eqref{eq.1.1} and Eq.~\eqref{eq.1.2}, provided
\begin{align}
\nonumber
&	a_1 = a_2 = (2-m)\beta^2, \quad
d_1 D^2 = 3 d_2 D^2 = -6B[m-(1-m)B]\beta^2, \\
\label{eq.11}
&	b_2 A^2 = 3 b_1 A^2 = -6(1+B) \beta^2.
\end{align}
Since $B > 0$, it follows that for this solution while $a_1, a_2 > 0$, $b_1, 
b_2 < 0$. On the other hand  $d_1, d_2 > (<)$ 0 depending on if $m < (> )$  
$(1-m)B$. Further, in case $b_2 = d_1$, then from Eq.~\eqref{eq.11} it follows that $d_2 = b_1$. 

On using the identities Eq.~\eqref{eq.6.1} and Eq.~\eqref{eq.6.2}, the coupled solution Eq.~\eqref{eq.11} can be rewritten as a superposition of the Jacobi elliptic functions 
\begin{subequations}
\begin{align}
\label{eq.12.1}
\phi_1(x) & = \frac{\beta}{\sqrt{2|b_1|}} [\dn(\beta x+\Delta,m)+\dn(\beta x-\Delta,m)], \\
\label{eq.12.2}
\phi_2(x) & =  \frac{\beta} {\sqrt{2 |d_2|}}[\dn(\beta x-\Delta,m)-\dn(\beta x+\Delta,m)],
\end{align}
\end{subequations}
where $B = \frac{m\sn^2(\Delta,m)}{\dn^2(\Delta,m)}$. The structure of these solutions is shown in Fig.~\ref{fig:Fig2} for three distinct choices of $m$ and a fixed value of $\Delta$.  In this and representative subsequent figures we will not show the variation with $\Delta$. \\ 

{\bf Solution III}

It is easy to check that
\begin{equation}\label{eq.13}
\phi_1(x) = \frac{A\cn(\beta x,m)}{1+B\cn^2(\beta x,m)}, \quad
\phi_2(x) = \frac{D\cn(\beta x,m) \sn(\beta x,m)}{1+B\cn^2(\beta x,m)}, \quad
\end{equation}
is an exact solution of the coupled equations Eq.~\eqref{eq.1.1}, and Eq.~\eqref{eq.1.2}, provided
\begin{align}
\nonumber
&	a_1 = [(2m-1) -6(1-m)B]\beta^2, \quad b_1 A^2 = 2(B+1)[3(1-m)B^2
+4(1-m)B-m]\beta^2, \\ 	
\nonumber
&	d_1 D^2 = -6[m+(1-m)B^2]B\beta^2, \\
\nonumber
&	a_2 = [(2m-1)-6(1-m)B]\beta^2, \quad b_2 A^2 = 6(B+1)[(1-m)B^2+2(1-m)B
-m]\beta^2, \\
\label{eq.14}
&       d_2 D^2 = -2[3(1-m)B^2+2(1-m)B+m]B\beta^2 \,. 
\end{align}
Notice that for this solution $d_1, d_2 < 0$ while $a_1 \ge (<)$ 0 depending on if $\frac{(2m-1)}{6(1-m)} \ge (<)$ $B$. On the other hand $a_2 \ge (<)$ 0 depending on whether $\frac{5m-4}{6(1-m)} \ge (<)$ $B$. Finally, $b_1 \ge (<)$ 0 depending on whether $(1-m)(B+1)(3B+1) \ge (<)$ 1 while $b_2 \ge (<)$ 1 depending on if $(1-m)(B+1)^2 \ge (<)$ 1. 

\begin{figure}[b]
\centering
\includegraphics[width=1\linewidth]{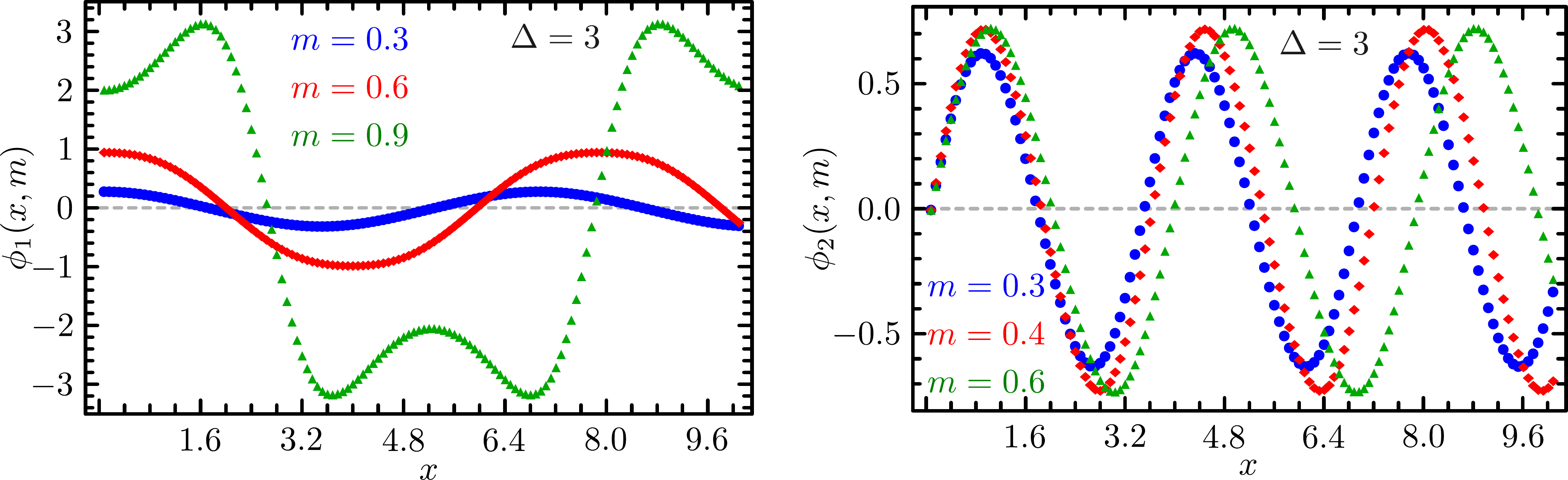} 
\caption{Variations of coupled fields $\phi_1(x,m)$ and $\phi_2(x,m)$ as a function of the position following Eq.~\eqref{eq.13} and Eq.~\eqref{eq.14}, respectively. Note that the scaling parameter $\beta$ is assumed to be 1, and $b_2 = d_1 = -1$. Since the $\phi_1$ amplitude is very large for $m=0.98$ we have instead used $m=0.9$.  Similarly, for $\phi_2$ we have kept the elliptic modulus range to be $0.3 < m < 0.6$.}\label{fig:Fig3}
\end{figure}

On using the identities Eq.~\eqref{eq.5.1} and Eq.~\eqref{eq.6.2}, the coupled solution Eq.~\eqref{eq.13} can be rewritten as 
\begin{subequations}
\begin{align}
\label{eq.15.1}
&	\phi_1(x) = \frac{A \dn^2(\Delta,m)}{2\cn(\Delta,m)} [\cn(\beta x+\Delta,)+\cn(\beta x-\Delta,m)], \\
\label{eq.15.2}
&	\phi_2(x) = \frac{D \dn^2(\Delta,m)}{2m\cn(\Delta,m)\sn(\Delta,m)} [\dn(\beta x-\Delta,m)-\dn(\beta x + \Delta,m)],
\end{align}
\end{subequations}
where $B = \frac{m\sn^2(\Delta,m)}{\dn^2(\Delta,m)}$ while $A$ and $D$ are as given by Eq.~\eqref{eq.14}. The structure of these solutions is shown in Fig.~\ref{fig:Fig3} for three distinct choices of $m$ and a fixed value of $\Delta$.  \\ 

{\bf Solution IV}

It is easy to check that
\begin{equation}\label{eq.16}
\phi_1(x) = \frac{A\dn(\beta x,m)}{1+B\cn^2(\beta x,m)}, \quad
\phi_2(x) = \frac{D \sn(\beta x,m)\dn(\beta x,m)}{1+B\cn^2(\beta x,m)}m, \quad 
\end{equation}
is an exact solution of the coupled equations Eq.~\eqref{eq.1.1} and Eq.~\eqref{eq.1.2}, provided
\begin{align}
\nonumber
&	[m-(1-m)B] a_1 = [(1-m)(4+m)B+m(2-m)]\beta^2, \\
\nonumber
&	[m-(1-m)B] b_1 A^2 = -2(B+1)[3(1-m)B^2+2(1-m)B+m]\beta^2, \\
\nonumber 
&	m[m-(1-m)B] d_1 D^2 = -6B[m- 2(1-m)B-(1-m)B^2]\beta^2, \\
\nonumber
&	[m-(1-m)B] a_2 = [(1-m)(4m+1)B+m(5-4m)]\beta^2, \\
\nonumber
&	m [m-(1-m)B] d_2 D^2 = -B[2m^2-5m(1-m)B -3(1-m^2) B^2]\beta^2, \\
\label{eq.17}
&	[m-(1-m)B] b_2 A^2  = -[6(1-m)B^3 +6(1-m)B^2+(m^2+4m+1)B+6m] \beta^2 \,. 
\end{align}
Notice that for this solution $a_1, a_2 > (<)$ 0 while $b_1, b_2 < (>)$ 0 depending on if $m > (<)$ $(1-m)B$. Finally, depending on the values of
$B$ and $m$, $d_1$ and $d_2$ can be positive or negative.

On using the identities Eq.~\eqref{eq.5.2} and Eq.~\eqref{eq.6.1}, the coupled solution Eq.~\eqref{eq.16} can be rewritten as 
\begin{subequations}
\begin{align}
\label{eq.18.1}
&	\phi_1(x) = \frac{A\dn(\Delta,m)}{2} [\dn(\beta x+\Delta,m)+\dn(\beta x-\Delta,m)], \\
\label{eq.18.2}
&	\phi_2(x) = \frac{D\dn(\Delta,m)}{2\sn(\Delta,m)} [\cn(\beta x-\Delta,m)-\cn(\beta x+\Delta,m)],
\end{align}
\end{subequations}
where $B = \frac{m\sn^2(\Delta,m)}{\dn^2(\Delta,m)}$ while $A$ and $D$ are as given by Eq.~\eqref{eq.17}. \\

{\bf Hyperbolic Limit}

In the limit $m = 1$, all four solutions, i.e. the solutions I to IV smoothly cross over to the hyperbolic limit~\cite{ks22a}
\begin{equation}\label{eq.19}
\phi_1(x) = \frac{A\cosh(\beta x)}{B+\cosh^2(\beta x)}, \quad
\phi_2(x) =  \frac{D \sinh(\beta x)} {B+\cosh^2(\beta x)}, \quad
\end{equation}
provided
\begin{align}
\nonumber
&	a_1 = a_2 = \beta^2\,,~~d_1 D^2 = 3 d_2 D^2 = -6B \beta^2, \\
\label{eq.20}
&	b_2 A^2 = 3 b_1 A^2 = -6(1+B) \beta^2.
\end{align}
On using Eq.~\eqref{eq.4}, it follows that for this solution $b_1, b_2, d_1, d_2 < 0$. Further, in case $b_2 = d_1$, then from Eq.~\eqref{eq.8} it follows that $d_2 = b_1$. 

On using the identities Eq.~\eqref{eq.6.1} and Eq.~\eqref{eq.6.2}, the coupled solution Eq.~\eqref{eq.19} can be rewritten as~\cite{ks22a} 
\begin{subequations}
\begin{align}
\label{eq.21.1}
&	\phi_1(x) = \frac{\beta}{\sqrt{2|b_1|}} [\sech(\beta x+\Delta)+\sech(\beta x-\Delta)], \\
\label{eq.21.2}
&	\phi_2(x) = \frac{\beta}{\sqrt{2|d_2|}} [\sech(\beta x-\Delta)-\sech(\beta x+\Delta)],
\end{align}
\end{subequations}
where $B = \sinh^2(\Delta)$. \\

{\bf Solution V}

It is readily checked that
\begin{equation}\label{eq.22}
\phi_1(x) = \frac{A\cn(\beta x,m)}{1+B\cn^2(\beta x,m)}, \quad
\phi_2(x) = \frac{D \sn(\beta x,m)}{1+B\cn^2(\beta x,m)}, \quad
\end{equation}
is an exact solution of the coupled equations Eq.~\eqref{eq.1.1} and Eq.~\eqref{eq.1.2}, provided
\begin{align}
\nonumber
&	B a_1 = [(2m-1)B+6m]\beta^2, \quad B d_1 D^2 = - 6[m+(1-m)B^2]\beta^2, \\
\nonumber
	&	B b_1 A^2 = - 2(B+1)[3m+4m B -(1-m)B^2]\beta^2, \\
\nonumber
&	B a_2 = [(5m-1)B+6m]\beta^2, \quad B d_2 D^2 = -2[3m+2m B +(1-m) B^2]\beta^2, \\
\label{eq.23}
&	B b_2 A^2  = - 6(B+1)[m+2mB-(1-m)B^2]\beta^2.
\end{align}
Notice that for this solution $d_1, d_2 < 0$ while $b_1 \ge (<)$ 0 provided $B^2 \ge (<)$ $m(B+1)(B+3)$. On the other hand, $b_2 \ge (<)$ 0 if $B^2 \ge (<)$ $m(B+1)^2$. Further, $a_1 \ge (<)$ 0 if $(2m-1)B+6m \ge (<)$ 0 while $a_2 \ge (<)$ 0 provided $(5m-1)B+6m \ge (<)$ 0.

On using the identities Eq.~\eqref{eq.3.1} and Eq.~\eqref{eq.5.1}, the coupled solution Eq.~\eqref{eq.22} can be rewritten as 
\begin{subequations}
\begin{align}
\label{eq.24.1}
&	\phi_1(x) = \frac{A \dn^2(\Delta,m)}{2\cn(\Delta,m)} [\cn(\beta x+\Delta,m)+\cn(\beta x-\Delta,m)], \\
\label{eq.24.2}
&	\phi_2(x) = \frac{D\dn(\Delta,m)}{2\cn(\Delta,m)} [\sn(\beta x+\Delta,m)+\sn(\beta x-\Delta,m)],
\end{align}
\end{subequations}
where $B = \frac{m\sn^2(\Delta,m)}{\dn^2(\Delta,m)}$ while $A$ and $D$ are as given by Eq.~\eqref{eq.23}. \\

{\bf Solution VI}

It is readily checked that
\begin{equation}\label{eq.25}
\phi_1(x) = \frac{A\dn(\beta x,m)}{1+B\cn^2(\beta x,m)}, \quad
\phi_2(x) = \frac{D \sn(\beta x,m)}{1+B\cn^2(\beta x,m)}, \quad
\end{equation}
is an exact solution of the coupled equations Eq.~\eqref{eq.1.1} and Eq.~\eqref{eq.1.2}, provided
\begin{align}
\nonumber
&	B a_1 = [(5m-4)B+6] \beta^2\,,~~B a_2 = [((5m-1)B+6m]\beta^2, \\
\nonumber
	&	B d_1 D^2 = -6[m-(1-m)B][m-2(1-m)B-(1-m)B^2] \beta^2, \\
\nonumber 
&	B b_1 A^2 = -2(B+1)[3m +2(3m-1)B-3(1-m)B^2]\beta^2, \\
\nonumber
	&	B d_2 D^2 = -2[m-(1-m)B][3m +2(3m-2)B -3(1-m)^2 B^2] \beta^2, \\
\label{eq.26}
&	B b_2 A^2  = -6(B+1)[m+2m B-(1-m)B^2]\beta^2.
\end{align}
Notice that for this solution $a_1 \ge (<)$ 0 depending on if $(5m-4)B+6 \ge (<)$ 0 while $a_2 \ge (<)$ 0 provided $(5m-1)B+6m \ge (<)$ 0. Similarly depending on the values of $B$ and $m$, $d_1, d_2, b_1, b_2$ can be positive or negative. 

On using the identities Eq.~\eqref{eq.3.1} and Eq.~\eqref{eq.6.1}, the coupled solution Eq.~\eqref{eq.25} can be rewritten as 
\begin{subequations}
\begin{align}
\label{eq.27.1}
&	\phi_1(x) = \frac{A\dn(\Delta)}{2} [\dn(\beta x+\Delta,m)+\dn(\beta x-\Delta,m)], \\
\label{eq.27.2}
&	\phi_2(x) = \frac{D \dn(\Delta,m)}{2 \cn(\Delta,m)} [\sn(\beta x+\Delta,m)+\sn(\beta x-\Delta,m)],
\end{align}
\end{subequations}
where $B = \frac{m\sn^2(\Delta,m)}{\dn^2(\Delta,m)}$ and $A$ and $D$ are as given by Eq.~\eqref{eq.26}. \\

{\bf Hyperbolic Limit}

In the limit $m = 1$, the solutions V and VI reduce to the hyperbolic superposed solution~\cite{ks22a}
\begin{equation}\label{eq.28}
\phi_1(x) = \frac{A\cosh(\beta x)}{B+\cosh^2(\beta x)}, \quad
\phi_2(x) = \frac{D \sinh(\beta x) \cosh(\beta x)}{B+\cosh^2(\beta x)}, \quad
\end{equation}
provided
\begin{align}
\nonumber
&	B a_1 = (B+6)\beta^2, \quad	B d_1 D^2 = - 6 \beta^2, \\
\nonumber
&	B b_1 A^2 = - 2(B+1)(4B+3) \beta^2, \\
\nonumber
&	B a_2 = 2(2B+3)\beta^2, \quad	B b_2 A^2  = - 6(B+1)(1+2B)\beta^2, \\
\label{eq.29}
&	B d_2 D^2 = -2(3+2 B)\beta^2.
\end{align}
Notice that for this solution $a_1, a_2 > 0$ while $b_1, b_2, d_1, d_2 < 0$. Further, in case $b_2 = d_1$, then from Eq.~\eqref{eq.29} it follows that $\frac{d_2}{b_1} = \frac{(2B+1)(2B+3)}{(4B+3)}$. 

On using the identities Eq.~\eqref{eq.3.1} and Eq.~\eqref{eq.5.1}, the coupled solution Eq.~\eqref{eq.28} can be re-expressed as~\cite{ks22a}
\begin{subequations}
\begin{align}
\label{eq.30.1}
&	\phi_1(x) = \frac{\sqrt{3\cosh(2\Delta)}\beta} {2\sqrt{|b_2|}} [\sech(\beta x+\Delta)+\sech(\beta x-\Delta)], \\
\label{eq.30.2}
&	\phi_2(x) = \frac{\sqrt{3}\beta}{\sqrt{2|d_1|}\sinh(\Delta)} [\tanh(\beta x+\Delta)+\tanh(\beta x-\Delta)],
\end{align}
\end{subequations}
where $B = \sinh^2(\Delta)$. \\

\begin{figure}[b]
\centering
\includegraphics[width=1\linewidth]{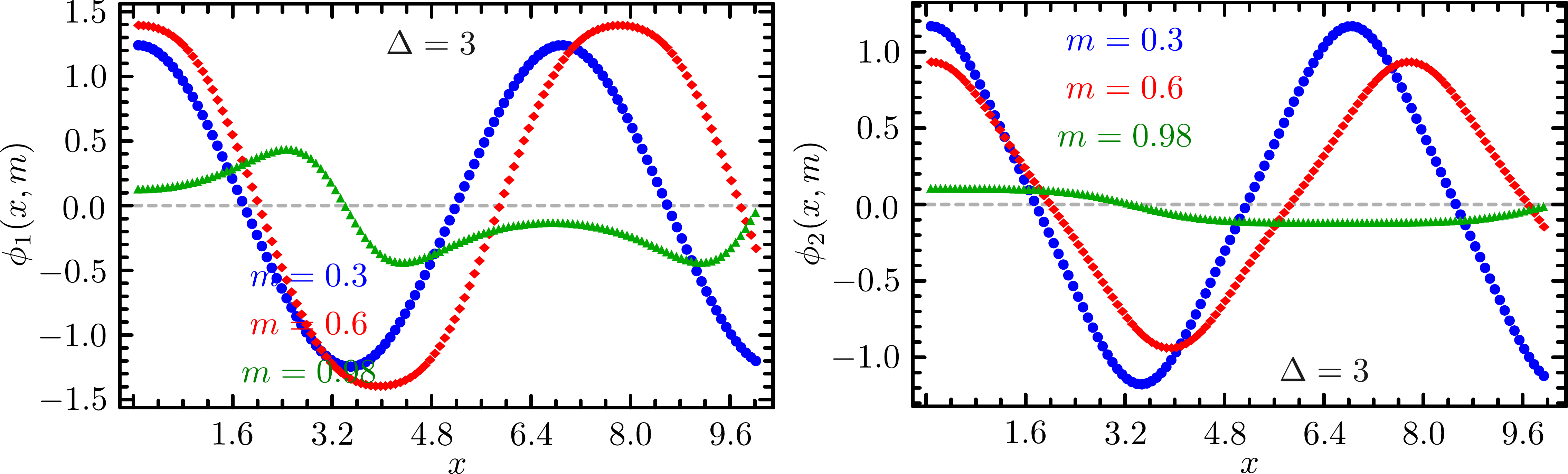} 
\caption{Variations of coupled fields $\phi_1(x,m)$ and $\phi_2(x,m)$ as a function of the position following Eq.~\eqref{eq.31} and Eq.~\eqref{eq.32}, respectively. Note that the scaling parameter $\beta$ is assumed to be 1, and $b_2 = d_1 = -1$.}\label{fig:Fig4}
\end{figure}

{\bf Solution VII}

It is readily checked that
\begin{equation}\label{eq.31}
\phi_1(x) = \frac{A\cn(\beta x,m)}{1+B\cn^2(\beta x,m)}, \quad
\phi_2(x) = \frac{D \cn(\beta x,m)\dn(\beta x,m)}{1+B\cn^2(\beta x,m)}, \quad
\end{equation}
is an exact solution of the coupled equations Eq.~\eqref{eq.1.1} and Eq.~\eqref{eq.1.2}, provided
\begin{align}
\nonumber
&	a_1 = [(2m-1)-6(1-m)B]\beta^2, \quad	m d_1 D^2 = 6B[m+(1-m)B^2]\beta^2, \\
\nonumber
	&	m b_1 A^2 = -2[m-(1-m)B][m+4m B-3(1-m)B^2]\beta^2, \\
\nonumber
&	a_2 = [(5m-1)-6(1-m)B]\beta^2, \quad	m d_2 D^2 = 2B[m-2m B +3(1-m) B^2]\beta^2, \\
\label{eq.32}
	&	m b_2 A^2  = -6[m-(1-m)B][m+2m B-(1-m)B^2]\beta^2.
\end{align}
Notice that for this solution $d_1 > 0$ while $d_2 \ge (<)$ 0 if $3B^2 \ge (<)$ $(B+1)(3B-1)m$. On the other hand, $a_1 \ge (<)$ 0 provided $(2m-1) \ge (<)$ $6(1-m)B$ while $a_2 \ge (<)$ 0 if $(5m-1) \ge (<)$ $6(1-m)B$. Finally, depending on the values of $B$ and $m$, both $b_1$ and $b_2$ can be positive or negative.

On using the identities Eq.~\eqref{eq.3.2} and Eq.~\eqref{eq.5.1}, the coupled solution Eq.~\eqref{eq.31} can be expressed differently as 
\begin{subequations}
\begin{align}
\label{eq.33.1}
&	\phi_1(x) = \frac{A \dn^{2}(\Delta)}{2 \cn(\Delta,m)} [\cn(\beta x+\Delta,m)+\cn(\beta x-\Delta,m)], \\
\label{eq.33.2}
&	\phi_2(x) = \frac{D \dn^2(\Delta,m)}{2\sn(\Delta,m)} [\sn(\beta x+\Delta,m)-\sn(\beta x-\Delta,m)],
\end{align}
\end{subequations}
where $B = \frac{m\sn^2(\Delta,m)}{\dn^2(\Delta,m)}$ while $A$ and $D$ are as given by Eq.~\eqref{eq.32}. The structure of these solutions is shown in Fig.~\ref{fig:Fig4} for three distinct choices of $m$ and a fixed value of $\Delta$.  \\

{\bf Solution VIII}

It is straightforward to check that
\begin{equation}\label{eq.34}
\phi_1(x) = \frac{A\dn(\beta x,m)}{1+B\cn^2(\beta x,m)}, \quad
\phi_2(x) = \frac{D \cn(\beta x,m) \dn(\beta x,m)}{1+B\cn^2(\beta x,m)}, \quad
\end{equation}
is an exact solution of the coupled equations Eq.~\eqref{eq.1.1} and Eq.~\eqref{eq.1.2}, provided
\begin{align}
\nonumber
&	[m-(1-m)B] a_1 = [m(2-m)+(1-m)(4+m)B]\beta^2, \\
\nonumber
&	[m-(1-m)B] b_1 A^2 = -2[m+2(1+m)B-(1-m)B^2] \beta^2, \\
\nonumber
&	[m-(1-m)B] d_1 D^2 = 6B[m-2(1-m)B-(1-m)B^2] B\beta^2, \\
\nonumber 
&	[m-(1-m)B] a_2 = [(5-m)m+(1-m^2)B]\beta^2, \\
\nonumber 
&	[m-(1-m)B] b_2 A^2 = -6[m+2m B-(1-m)B^2]\beta^2, \\
\label{eq.35}
&	[m-(1-m)B] d_2 D^2 = 2B [m-2(2-m)B-(1-m)B^2]\beta^2.
\end{align}
Notice that for this solution while $a_1, a_2 > (<)$ 0 depending on if $m > (<)$ $(1-m)B$, depending on the values of $B$ and $m$, $b_1, b_2, d_1, d_2$ can take positive or negative values.

On using the identities Eq.~\eqref{eq.3.2} and Eq.~\eqref{eq.6.1}, the coupled solution Eq.~\eqref{eq.34} can be expressed differently as 
\begin{subequations}
\begin{align}
\label{eq.36.1}
&	\phi_1(x) = \frac{A\dn(\Delta,m)}{2} [\dn(\beta x+\Delta,m)+\dn(\beta x-\Delta,m)], \\
\label{eq.36.2}
&	\phi_2(x) = \frac{D\dn^2(\Delta,m)}{2\sn(\Delta,m)} [\sn(\beta x+\Delta,m)-\sn(\beta x-\Delta,m)],
\end{align}
\end{subequations}
where $B = \frac{m\sn^2(\Delta,m)}{\dn^2(\Delta,m)}$ while $A$ and $D$ are as given by Eq.~\eqref{eq.35}. \\ 

{\bf Hyperbolic Limit}

In the limit $m = 1$, the solutions VII and VIII become the well-known hyperbolic solution~\cite{ks22a}
\begin{equation}\label{eq.37}
\phi_1(x) = \frac{A\cosh(\beta x)}{B+\cosh^2(\beta x)}, \quad
\phi_2(x) = \frac{D}{B+\cosh^2(\beta x)}, \quad
\end{equation}
provided
\begin{align}
\nonumber
&	a_1 = \beta^2\,,~~d_1 D^2 = 6B \beta^2, \quad	b_1 A^2 = -2(1+4B)\beta^2, \\
\label{eq.38}
&	a_2 = 4\beta^2\,,~~d_2 D^2 = 2B(1-2 B)\beta^2, \quad	b_2 A^2 = -6(1+2B)\beta^2.
\end{align}
Notice that for this solution while $a_1, a_2 > 0$, $b_1,b_2 < 0$, and $d_1 > 0$, while $d_2 > (<)$ 0 depending on if $B < (>)$ 1/2. 

On using the identities Eq.~\eqref{eq.3.2} and Eq.~\eqref{eq.5.1}, the coupled solution Eq.~\eqref{eq.37} can be rewritten as~\cite{ks22a}
\begin{subequations}
\begin{align}
\label{eq.39.1}
&	\phi_1(x) = \frac{\sqrt{3\cosh(2\Delta)}\beta} {2\sqrt{|b_2|}\cosh(\Delta)} [\sech(\beta x+\Delta)+\sech(\beta x-\Delta)], \\
\label{eq.39.2}
&	\phi_2(x) = \frac{\sqrt{3}\beta}{\sqrt{2|d_1|}\sech(\Delta)} [\tanh(\beta x+\Delta)-\tanh(\beta x-\Delta)],
\end{align}
\end{subequations}
where $B = \sinh^2(\Delta)$. \\

{\bf Solution IX}

It is easy to check that
\begin{equation}\label{eq.40}
\phi_1(x) = \frac{A\cn(\beta x,m)\sn(\beta x,m)}{1+B\cn^2(\beta x,m)}, \quad
\phi_2(x) = \frac{D \cn(\beta x,m)\dn(\beta x,m)}{1+B\cn^2(\beta x,m)}, \quad
\end{equation}
is an exact solution of the coupled equations Eq.~\eqref{eq.1.1} and Eq.~\eqref{eq.1.2}, provided~\cite{ks22a}
\begin{align}
\nonumber
&	a_1 = [(5m-4)-6(1-m)B]\beta^2, \quad	a_2 = [((5m-1)-6(1-m)B]\beta^2, \\
\nonumber
&	d_1 D^2 = -6(B+1)[m-2(1-m) B -(1-m)B^2]\beta^2, \\
\nonumber
	&	b_1 A^2 = -2[m-(1-m)B][3m +2(3m-1) B-3(1-m) B^2]\beta^2, \\
\nonumber
&	d_2 D^2 = -2m(B+1)[3+2(3m-2)B -3(1-m)B^2]\beta^2, \\
\label{eq.41}
	&	b_2 A^2  = -6[m-(1-m)B][m+2m B-(1-m) B^2]\beta^2.
\end{align}
Notice that for this solution while $a_1 > (<)$ 0 depending on if $5m-4 \ge (<)$ $6(1-m)B$, $a_2 \ge (<)$ 0 depending on whether $5m-1 \ge (<)$ $6(1-m)B$. Further, depending on the values of $B$ and $m$, $b_1, b_2, d_1, d_2$ can take positive or negative values.

\begin{figure}[t]
\centering
\includegraphics[width=1\linewidth]{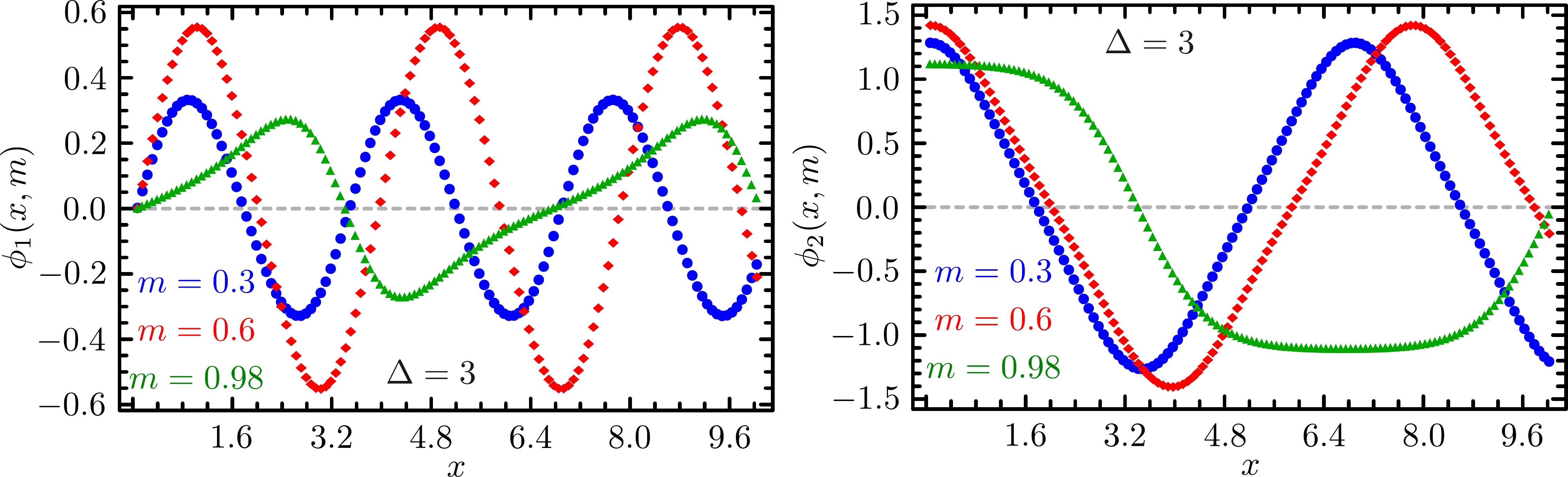} 
\caption{Variations of coupled fields $\phi_1(x,m)$ and $\phi_2(x,m)$ as a function of the position following Eq.~\eqref{eq.40} and Eq.~\eqref{eq.41}, respectively. Note that the scaling parameter $\beta$ is assumed to be 1, and $b_1 = d_2 = -1$.}\label{fig:Fig5}
\end{figure}

On using the identities Eq. \eqref{eq.3.2} and Eq.~\eqref{eq.6.2}, the coupled solution Eq.~\eqref{eq.40} can be rewritten as~\cite{ks22a}
\begin{subequations}
\begin{align}
\label{eq.42.1}
&	\phi_1(x) = \frac{A\dn^2(\Delta,m)}{2m \sn(\Delta,m)\cn(\Delta,m)} [\dn(\beta x-\Delta,m)-\dn(\beta x+\Delta,m)], \\
\label{eq.42.2}
&	\phi_2(x) = \frac{D \dn^2(\Delta,m)}{2\sn(\Delta,m)} [\sn(\beta x+\Delta,m)-\sn(\beta x-\Delta,m)],
\end{align}
\end{subequations}
where $B = \frac{m\sn^2(\Delta,m)}{\dn^2(\Delta,m)}$ while $A$ and $D$ are as given by Eq.~\eqref{eq.41}. The structure of these solutions is shown in Fig.~\ref{fig:Fig5} for three distinct choices of $m$ and a fixed value of $\Delta$. \\

{\bf Solution X}

It is readily checked that
\begin{equation}\label{eq.43}
\phi_1(x) = \frac{A\sn(\beta x,m)\dn(\beta x,m)}{1+B\cn^2(\beta x,m)}, \quad
\phi_2(x) = \frac{D \cn(\beta x,m) \dn(\beta x,m)} {1+B\cn^2(\beta x,m)}, \quad
\end{equation}
is an exact solution of the coupled equations Eq.~\eqref{eq.1.1} and Eq.~\eqref{eq.1.2}, provided
\begin{align}
\nonumber
&	[m-(1-m)B] a_1 = [(1-m)(1+4m)B+m(5-4m)]\beta^2, \\
\nonumber
&	[m-(1-m)B] b_1 A^2 = -2[3m +4m B -(1-m) B^2]\beta^2, \\
\nonumber
	&	[m-(1-m)B] d_1 D^2 = -[6m+2m(4-m)B+2(1-m)(2+3m)B^2+2(1-m)
(2m+1)B^3]\beta^2, \\
\nonumber 
&	[m-(1-m)B] a_2 = [(5-m)m +(1-m^2)B]\beta^2, \\
\nonumber
&	[m- (1-m)B] b_2 A^2 = -6 [m +2mB-(1-m)B^2]\beta^2, \quad 0 < m < 1, \\
\label{eq.44}
&	[m-(1-m)B] d_2 D^2 = -[6m +10mB+2(1+m) B^2+(1-m)B^3]\beta^2.
\end{align}
Notice that for this solution while $a_1, a_2 > (<)$ 0, $d_1, d_2 < (>)$ 0 depending on if $m > (<)$ $(1-m)B$. Further, depending on the values of $B$ and $m$, both $b_1, b_2$ can take positive or negative values.

On using the identities Eq.~\eqref{eq.3.2} and Eq.~\eqref{eq.5.2}, the coupled solution Eq.~\eqref{eq.43} can be rewritten as 
\begin{subequations}
\begin{align}
\label{eq.45.1}
&	\phi_1(x) = \frac{A\dn(\Delta,m)}{2\sn(\Delta,m)} [\cn(\beta x-\Delta,m)-\cn(\beta x+\Delta,m)], \\
\label{eq.45.2}
&	\phi_2(x) = \frac{D\dn^2(\Delta,m)}{2\sn(\Delta,m)} [\sn(\beta x+\Delta,m)-\sn(\beta x-\Delta,m)],
\end{align}
\end{subequations}
where $B = \frac{m\sn^2(\Delta,m)}{\dn^2(\Delta,m)}$ while $A$ and $D$ are as given by Eq.~\eqref{eq.44}. \\

{\bf Hyperbolic Limit}

In the limit $m = 1$, the solutions IX and X go over to the hyperbolic solution~\cite{ks22a}
\begin{equation}\label{eq.46}
\phi_1(x) = \frac{A\sinh(\beta x)}{B+\cosh^2(\beta x)}, \quad
\phi_2(x) = \frac{D}{B+\cosh^2(\beta x)}, \quad
A,B,D > 0,
\end{equation}
provided
\begin{align}
\nonumber
&	a_1 = \beta^2\,,~~d_1 D^2 = -6(B+1)\beta^2, \quad	b_1 A^2 = -2(3+4B)\beta^2, \\
\label{eq.47}
&	a_2 = 4\beta^2\,,~~d_2 D^2 = -2(B+1)(3+2B)\beta^2, \quad	b_2 A^2  = -6(1+2B)\beta^2.
\end{align}
Notice that for this solution while $a_1, a_2 > 0$, $b_1, b_2, d_1, d_2 < 0$. Further, in case $b_2 = d_1$, then it follows from Eq.~\eqref{eq.47} that $\frac{d_2}{b_1} = \frac{(3+2B)(1+2B)}{3+4B}$. 

On using the identities Eq.~\eqref{eq.3.2} and Eq.~\eqref{eq.6.2}, the coupled solution Eq.~\eqref{eq.46} can be rewritten as~\cite{ks22a}
\begin{subequations}
\begin{align}
\label{eq.48.1}
&	\phi_1(x) = \frac{\sqrt{3\cosh(2\Delta)}\beta} {2\sqrt{|b_2|}\sech(\Delta)} [\sech(\beta x-\Delta)-\sech(\beta x+\Delta)], \\
\label{eq.48.2}
&	\phi_2(x) = \frac{\sqrt{3}\beta}{\sqrt{2|d_1|}\sinh(\Delta)} [\tanh(\beta x+\Delta)-\tanh(\beta x-\Delta)], 
\end{align}
\end{subequations}
where $B = \sinh^2(\Delta)$. \\

{\bf Solution XI}

\begin{figure}[b]
\centering
\includegraphics[width=1\linewidth]{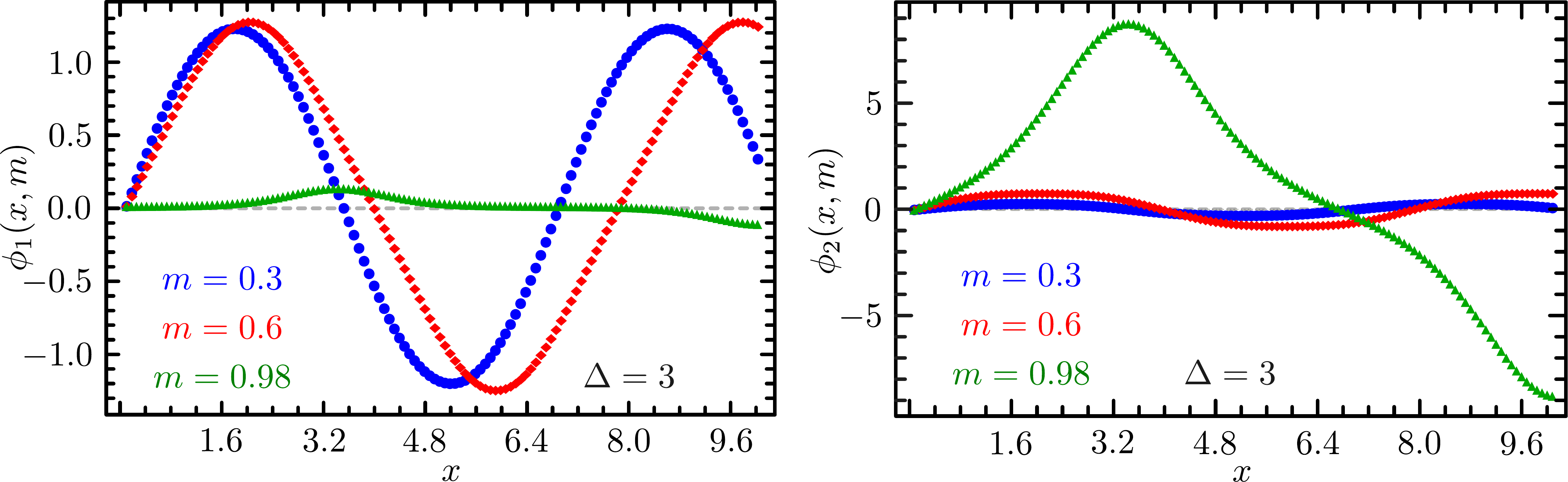} 
\caption{Variations of coupled fields $\phi_1(x,m)$ and $\phi_2(x,m)$ as a function of the position following Eq.~\eqref{eq.49} and Eq.~\eqref{eq.50}, respectively. Note that the scaling parameter $\beta$ is assumed to be 1, and $b_2 = 1, d_1 = -1$.}\label{fig:Fig6}
\end{figure}

It is straightforward to check that
\begin{equation}\label{eq.49}
\phi_1(x) = \frac{A\sn(\beta x,m)}{1+B\cn^2(\beta x,m)}, \quad
\phi_2(x) = \frac{D \sn(\beta x,m) \dn(\beta x,m)} {1+B\cn^2(\beta x,m)}, \quad
\end{equation}
is an exact solution of the coupled equations Eq.~\eqref{eq.1.1} and Eq.~\eqref{eq.1.2}, provided
\begin{align}
\nonumber
&	(1+B) a_1 = [(5-m)B -(1+m)]\beta^2, \quad	(1+B) a_2 = [(5-4m)B-(4m+1)]\beta^2, \\
\nonumber
&	m (1+B) d_1 D^2 = -6B [m +2mB-(1-m)B^2]\beta^2, \\
\nonumber 
	&	m(1+B) b_1 A^2 = 2[m-(1-m)B][m -2m B+3(1-m)B^2]\beta^2, \\
\nonumber 
&	m (1+B) d_2 D^2 = -2[m +4m B -3(1-m) B^2] B \beta^2, \\
\label{eq.50}
	&	m(1+B) b_2 A^2  = 6[m-(1-m)B][m +(1-m)B^2]\beta^2.
\end{align}
Notice that for this solution while $b_2 > (<)$ 0 depending on if $m > (<)$  $(1-m)B$, $a_1 > (<)$ 0 depending on if $(5-m)B \ge (<)$ $(1+m)$, $a_2 \ge (<)$ 0 depending on whether $(5-4m)B \ge (<)$ $4m+1$. Further, depending on the values of $B$ and $m$, $b_1, d_1, d_2$
can take positive or negative values.

On using the identities Eq.\eqref{eq.3.1} and Eq.~\eqref{eq.5.2}, the coupled solution Eq.\eqref{eq.49} can be rewritten as 
\begin{subequations}
\begin{align}
\label{eq.51.1}
&	\phi_1(x) = \frac{A\dn(\Delta,m)}{2 \cn(\Delta,m)} [\sn(\beta x+\Delta,m)+\sn(\beta x-\Delta,m)], \\
\label{eq.51.2}
&	\phi_2(x) = \frac{D\dn(\Delta,m)}{2\sn(\Delta,m)} [\cn(\beta x-\Delta,m)-\cn(\beta x+\Delta,m)],
\end{align}
\end{subequations}
where $B = \frac{m\sn^2(\Delta,m)}{\dn^2(\Delta,m)}$ while $A$ and $D$ are as given by Eq.~\eqref{eq.50}. The structure of these solutions is shown in Fig.~\ref{fig:Fig6} for three distinct choices of $m$ and a fixed value of $\Delta$. \\ 

{\bf Solution XII}

\begin{figure}[b]
\centering
\includegraphics[width=1\linewidth]{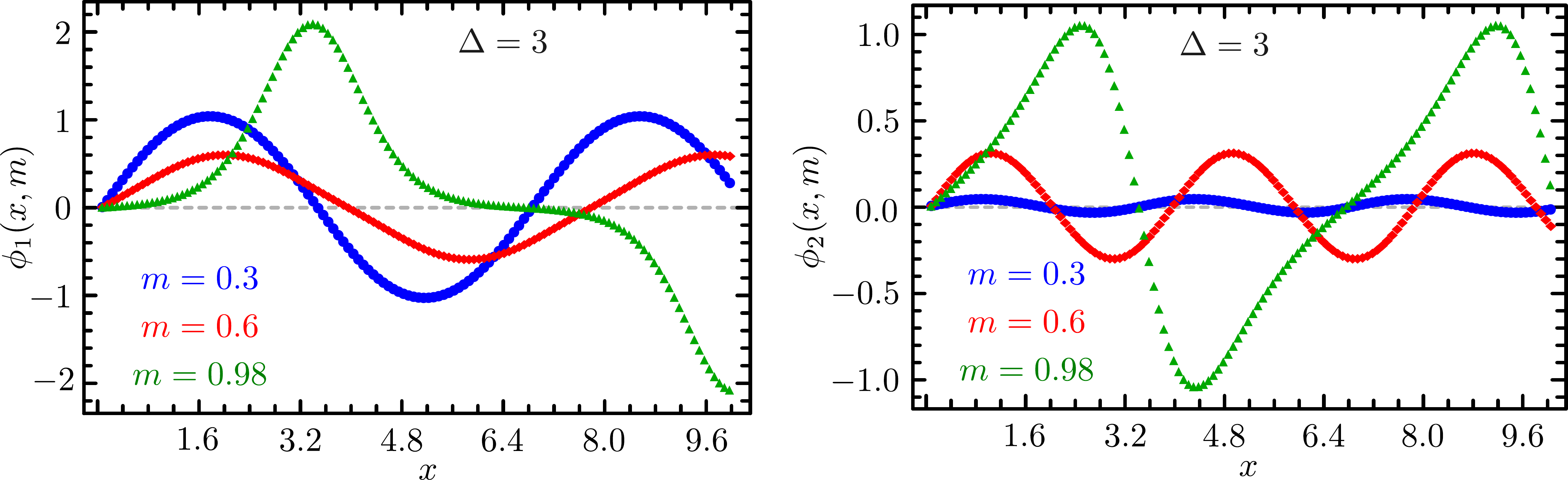} 
\caption{Variations of coupled fields $\phi_1(x,m)$ and $\phi_2(x,m)$ as a function of the position following Eq.~\eqref{eq.52} and Eq.~\eqref{eq.53}, respectively. Note that the scaling parameter $\beta$ is assumed to be 1, and $b_2 = d_1 = -1$.}\label{fig:Fig7}
\end{figure}

It is easy to check that
\begin{equation}\label{eq.52}
\phi_1(x) = \frac{A\sn(\beta x,m)}{1+B\cn^2(\beta x,m)}, \quad
\phi_2(x) = \frac{D \sn(\beta x,m) \cn(\beta x,m)} {1+B\cn^2(\beta x,m)}, \quad
\end{equation}
is an exact solution of the coupled equations Eq.~\eqref{eq.1.1} and Eq.~\eqref{eq.1.2}, provided
\begin{align}
\nonumber
&	(1+B) a_1 = [(5-m)B-(1+m)]\beta^2, \quad	(1+B) a_2 = [(2-m)B-(4+m)]\beta^2, \\
\nonumber
&	(1+B) d_1 D^2 = -6B [m+2mB-(1-m)B^2]\beta^2, \\
\nonumber
&	(1+B) b_1 A^2 = 2[m-2(2-m)B-(1-m)B^2] \beta^2, \\
\nonumber
&	(1+B) d_2 D^2 = -2B [m+2(1+m)B-(1-m)B^2] \beta^2, \\
\label{eq.53}
&	(1+B) b_2 A^2 = 6[m-2(1-m)B-(1-m)B^2]\beta^2.
\end{align}
Notice that for this solution while $a_1 > (<)$ 0 depending on if $(5-m)B \ge (<)$ $(1+m)$, $a_2 \ge (<)$ 0 depending on whether $(2-m)B \ge (<)$ $4+m$. Further, depending on the values of $B$ and $m$, $b_1, b_2, d_1, d_2$ can take positive or negative values.

On using the identities Eq.~\eqref{eq.3.1} and Eq.~\eqref{eq.6.2}, the coupled solution Eq.~\eqref{eq.52} can be rewritten as 
\begin{subequations}
\begin{align}
\label{eq.54.1}
&	\phi_1(x) = \frac{A\dn(\Delta,m)}{2\cn(\Delta,m)} [\sn(\beta x+\Delta,m)+\sn(\beta x-\Delta,m)], \\
\label{eq.54.2}
&	\phi_2(x) = \frac{D\dn^2(\Delta,m)}{2\cn(\Delta,m)\sn(\Delta,m)} [\dn(\beta x-\Delta,m)-\dn(\beta x+\Delta,m)],
\end{align}
\end{subequations}
where $B = \frac{m\sn^2(\Delta,m)}{\dn^2(\Delta,m)}$ while $A$ and $D$ are as given by Eq.~\eqref{eq.53}. The structure of these solutions is shown in Fig.~\ref{fig:Fig7} for three distinct choices of $m$ and a fixed value of $\Delta$. \\ 

{\bf Hyperbolic Limit}

In the limit $m = 1$, the solutions XI and XII go over to the hyperbolic solution~\cite{ks22a}
\begin{equation}\label{eq.55}
\phi_1(x) = \frac{A\sinh(\beta x)\cosh(\beta x)}{B+\cosh^2(\beta x)}, \quad
\phi_2(x) = \frac{D \sinh(\beta x)}{B+\cosh^2(\beta x)}, \quad
\end{equation}
provided
\begin{align}
\nonumber
&	(1+B) a_1 = 2(2B-1)\beta^2, \quad	(1+B) d_1 D^2= -6B (1+2B)\beta^2, \\
\nonumber
&	(1+B) b_1 A^2 = 2(1-2B) \beta^2, \quad	(1+B) a_2 = (B-5)\beta^2, \\
\label{eq.56}
&	(1+B) d_2 D^2 = -2B (1+4B) \beta^2, \quad	(1+B) b_2 A^2 = 6\beta^2.
\end{align}
Notice that for this solution whereas $b_2 > 0$, $d_1, d_2 < 0$, while other three parameters, i.e. $a_1, a_2, b_1$ depend on the value of $B$. In particular, in case $B > (<)$ 1/2 then $a_1 > (<)$ 0 while $b_1 < (>)$ 0. On the other hand $a_2 > (<)$ 0 depending on if $B > ( < )$ 5. Thus such a solution does not exist in case $d_1 = b_2$.

On using the identities Eq.~\eqref{eq.3.1} and Eq.~\eqref{eq.6.2}, the coupled solution Eq.~\eqref{eq.55} can be rewritten as~\cite{ks22a}
\begin{subequations}
\begin{align}
\label{eq.57.1}
&	\phi_1(x) = \frac{\sqrt{3}\beta}{\sqrt{2 b_2}\cosh(\Delta)} [\tanh(\beta x+\Delta)+\tanh(\beta x-\Delta)], \\
\label{eq.57.2}
&	\phi_2(x) = \frac{\sqrt{3\cosh(2\Delta)}\beta}{\sqrt{2|d_1|}\cosh(\Delta)} [\sech(\beta x-\Delta)-\sech(\beta x+\Delta)],
\end{align}
\end{subequations}
where $B = \sinh^2(\Delta)$.  \\

{\bf Solution XIII}

It is straightforward to check that
\begin{equation}\label{eq.58}
\phi_1(x) = \frac{A\sn(\beta x,m)}{1+B\cn^2(\beta x,m)}, \quad
\phi_2(x) = \frac{D \cn(\beta x,m) \dn(\beta x,m)} {1+B\cn^2(\beta x,m)}, \quad
\end{equation}
is an exact solution of the coupled equations Eq.~\eqref{eq.1.1} and Eq.~\eqref{eq.1.2}, provided
\begin{align}
\nonumber
&	a_1 = a_2 = -(1+m)\beta^2, \quad	d_1 D^2 = 3d_2 D^2 = 6B(B+1)\beta^2, \\
\label{eq.59}
&	b_2 A^2 = 3 b_1 A^2 = 6[m-(1-m)B] \beta^2 .
\end{align}
Thus for such a solution while $a_1, a_2 < 0$, $d_1, d_2 > 0$. On the other hand, $b_1, b_2 > (<)$ 0 depending on whether $m > (<)$ $(1-m)B$.

\begin{figure}[t]
\centering
\includegraphics[width=1\linewidth]{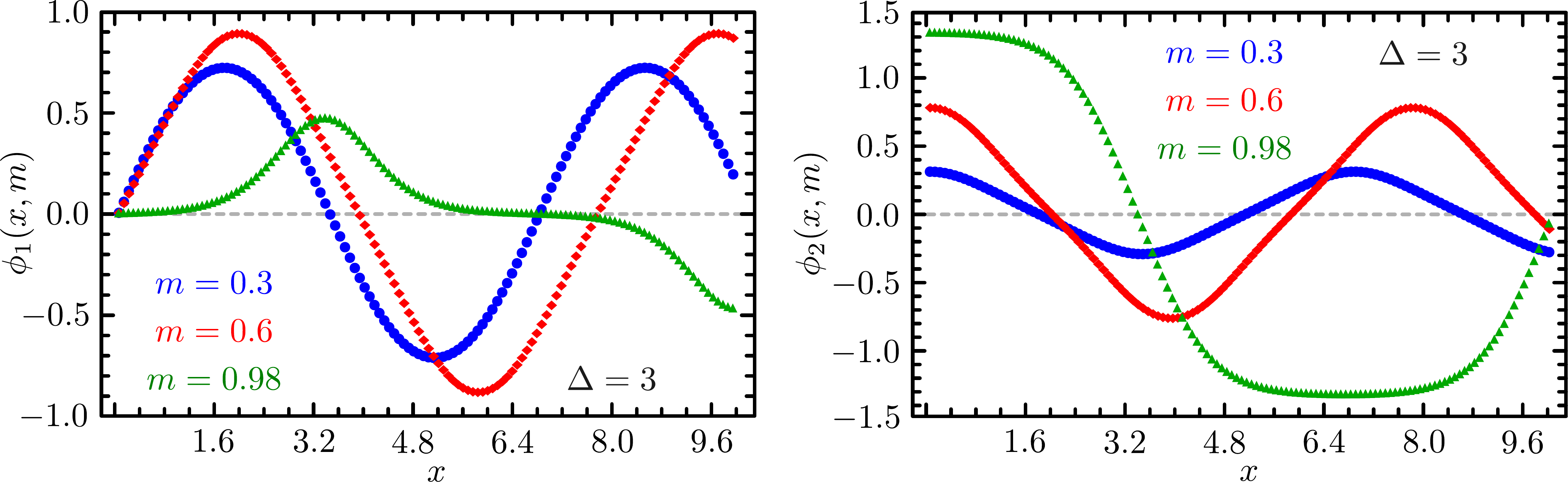} 
\caption{Variations of coupled fields $\phi_1(x,m)$ and $\phi_2(x,m)$ as a function of the position following Eq.~\eqref{eq.58} and Eq.~\eqref{eq.59}, respectively. Note that the scaling parameter $\beta$ is assumed to be 1, and $b_2 = d_1 = -1$.}\label{fig:Fig8}
\end{figure}

On using the identities Eq.\eqref{eq.3.1} and Eq.~\eqref{eq.3.2}, the coupled solution Eq.~\eqref{eq.58} can be rewritten as 
\begin{subequations}
\begin{align}
\label{eq.60.1}
&	\phi_1(x) =  \dn(\Delta,m)\frac{\sqrt{m} \beta}{\sqrt{2 b_1}} [\sn(\beta x+\Delta,m)+\sn(\beta x-\Delta,m)], \\
\label{eq.60.2}
&	\phi_2(x) = \frac{\sqrt{m}\beta} {\sqrt{2d_2}}[\sn(\beta x+\Delta,m)-\sn(\beta x-\Delta,m)],
\end{align}
\end{subequations}
where $B = \frac{m\sn^2(\Delta,m)}{\dn^2(\Delta,m)}$. The structure of these solutions is shown in Fig.~\ref{fig:Fig8} for three distinct choices of $m$ and a fixed value of $\Delta$. \\ 

{\bf Hyperbolic Limit}

In the limit $m = 1$, the solution XIII goes over to the hyperbolic solution~\cite{ks22a}
\begin{equation}\label{eq.61}
\phi_1(x) = \frac{A\sinh(\beta x)\cosh(\beta x)}{B+\cosh^2(\beta x)}, \quad
\phi_2(x) = \frac{D}{B+\cosh^2(\beta x)}, \quad
A,B,D > 0,
\end{equation}
provided
\begin{align}
\nonumber
&	a_1 = a_2 = -2\beta^2\,,~~d_1 D^2 = 3d_2 D^2 = 6B(B+1)\beta^2, \\
\label{eq.62}
&	b_2 A^2 = 3 b_1 A^2 = 6\beta^2.
\end{align}
Thus for such a solution while $a_1, a_2 < 0$, $b_1, b_2, d_1, d_2 > 0$. Further, if $d_1 = b_2$, Eq.~\eqref{eq.62} implies that $d_2 = b_1$. 

On using the identities Eq.~\eqref{eq.3.1} and Eq.~\eqref{eq.3.2}, the coupled solution Eq.~\eqref{eq.61} can be rewritten as~\cite{ks22a}
\begin{subequations}
\begin{align}
\label{eq.63.1}
&	\phi_1(x) = \frac{\beta}{\sqrt{2 b_1}} [\tanh(\beta x+\Delta)+\tanh(\beta x-\Delta)], \\
\label{eq.63.2}
&	\phi_2(x) = \frac{\beta}{\sqrt{2d_2}} [\tanh(\beta x+\Delta)-\tanh(\beta x-\Delta)],
\end{align}
\end{subequations}
where $B = \sinh^2(\Delta)$. \\

{\bf Solution XIV}

It is easy to check that
\begin{equation}\label{eq.64}
\phi_1(x) = \frac{A\cn(\beta x,m)}{1+B\cn^2(\beta x,m)}, \quad
\phi_2(x) = \frac{D\dn(\beta x,m)}{1+B\cn^2(\beta x,m)}, \quad
\end{equation}
is an exact solution of the coupled equations Eq.~\eqref{eq.1.1} and Eq.~\eqref{eq.1.2}, provided
\begin{align}
\nonumber
&	B a_1 = [(2m-1)B + 6m]\beta^2, \quad	(1-m)B d_1 D^2 = -6[(1-m)B^2 
+m]\beta^2, \\
\nonumber
&       (1-m)B b_1 A^2 = 2[m-(1-m)B][3m-4(1-m)B-(1-m)B^2]\beta^2, \\
\nonumber
&	B a_2 = [(5m-4)B+6m]\beta^2, \quad	(1-m)B d_2 D^2 = 2[2(1-m)B 
-(1-m)B^2-3m]\beta^2, \\
\label{eq.65}
&	(1-m)B b_2 A^2 = 6[2m^2-3m(1-m)B +(1-m)(2-3m)B^2+(1-m)^2 B^3] \beta^2.
\end{align}
Notice that for this solution $0 < m < 1$ while $d_1 < 0$. Further, $a_1 \ge (<)$ 0 depending on if $2m(B+3) \ge (<)$ $B$, $a_2 \ge (<)$ 0 depending on whether $m(5B+6) \ge (<)$  $4B$. Finally, depending on the values of $B$ and $m$, $b_1, b_2, d_2$ can take positive or negative values.

On using the identities Eq.~\eqref{eq.5.1} and Eq.~\eqref{eq.6.1}, the coupled solution Eq.~\eqref{eq.64} can be expressed differently as 
\begin{subequations}
\begin{align}
\label{eq.66.1}
&	\phi_1(x) = \frac{A\dn^2(\Delta,m)}{2\cn(\Delta,m)} [\cn(\beta x+\Delta,m)+\cn(\beta x-\Delta,m)], \\
\label{eq.66.2}
&	\phi_2(x) = \frac{D\dn(\Delta,m)}{2} [\dn(\beta x+\Delta,m)+\dn(\beta x-\Delta,m)],
\end{align}
\end{subequations}
where $B = \frac{m\sn^2(\Delta,m)}{\dn^2(\Delta,m)}$ while $A$ and $D$ are as given by Eq.~\eqref{eq.65}.  \\

{\bf Solution XV}

It is easy to check that
\begin{equation}\label{eq.67}
\phi_1(x) = \frac{A\cn(\beta x,m) \sn(\beta x,m)}{1+B\cn^2(\beta x,m)}, \quad
\phi_2(x) = \frac{D \sn(\beta x,m)\dn(\beta x,m)}{1+B\cn^2(\beta x,m)}, \quad
\end{equation}

is an exact solution of the coupled equations Eq.~\eqref{eq.1.1} and Eq.~\eqref{eq.1.2}, provided
\begin{align}
\nonumber
&	(B+1) a_1 = [(2-m)B-(4+m)]\beta^2, \quad	(B+1) a_2 = [(5-4m)B-(4m+1)]\beta^2, \\
\nonumber
&	(1-m)(B+1) d_1 D^2 = 6[1- (1-m)(1+B)^2]\beta^2, \\
\nonumber
&	(1-m)(B+1) b_1 A^2 = -2[m-(1-m)^2 B][3m-2(1-m)B+(1-m)^2 B^2]\beta^2, \\
\nonumber
&	(1-m) (B+1) b_2 A^2 = -6[m-(1-m)B][m+(1-m)B^2] \beta^2, \\
\label{eq.68}
&	(1-m)(B+1) d_2 D^2 = 2[3m-4(1-m)B -(1-m) B^2]\beta^2.
\end{align}
Notice that for this solution $0 < m < 1$ while $b_2 < (>)$ 0 depending on if $m > (<)$ $(1-m)B$. Further, $a_1 \ge (<)$ 0 depending on if
$(2-m)B) \ge (<)$ $4+m$, $a_2 \ge (<)$ 0 depending on whether $(5-4m)B \ge (<)$ $4m+1$. Finally, depending on the values of $B$ and $m$, $b_1, d_1, d_2$
can take positive or negative values.

\begin{figure}[t]
\centering
\includegraphics[width=1\linewidth]{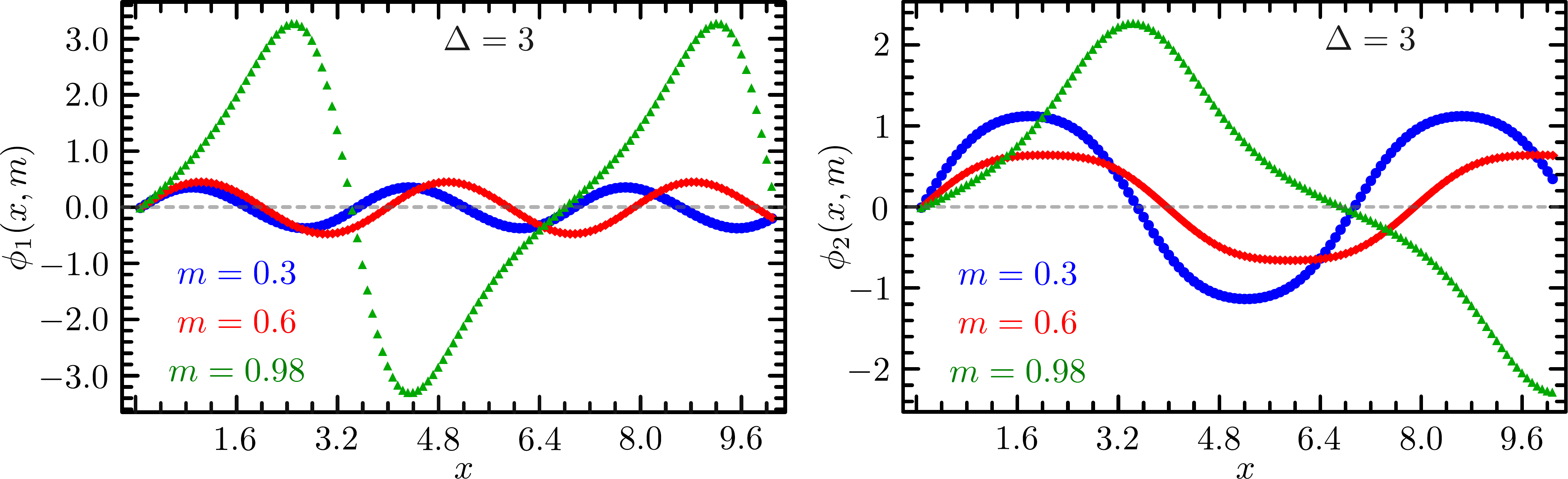} 
\caption{Variations of coupled fields $\phi_1(x,m)$ and $\phi_2(x,m)$ as a function of the position following Eq.~\eqref{eq.67} and Eq.~\eqref{eq.68}, respectively. Note that the scaling parameter $\beta$ is assumed to be 1, and $b_2 = d_1 = -1$.}\label{fig:Fig9}
\end{figure}

On using the identities Eq.~\eqref{eq.5.2} and Eq.~\eqref{eq.6.2}, the coupled solution Eq.~\eqref{eq.67} can be expressed differently as 
\begin{subequations}
\begin{align}
\label{eq.69.1}
&	\phi_1(x) = \frac{A \dn^2(\Delta,m)}{2m\sn(\Delta,m) \cn(\Delta,m)} [\dn(\beta x-\Delta,m)-\dn(\beta x+\Delta,m)], \\
\label{eq.69.2}
&	\phi_2(x) = \frac{D \dn(\Delta,m)}{2\sn(\Delta,m)} [\cn(\beta x-\Delta,m)-\cn(\beta x+\Delta,m)],
\end{align}
\end{subequations}
where $B = \frac{m\sn^2(\Delta,m)}{\dn^2(\Delta,m)}$ while $A$ and $D$ are as given by Eq.~\eqref{eq.68}. The structure of these solutions is shown in Fig.~\ref{fig:Fig9} for three distinct choices of $m$ and a fixed value of $\Delta$. 

\section{The Periodic But Not Superposed Solutions When $B = m/(1-m)$ and when $m = 0$}

We now discuss the fifteen solutions obtained in the previous section in case $m = (1-m)B$ as well as when $m = 0$. We will show that in both these cases while we have exact solutions but there are {\it no} superposed solutions. Let us discuss the two cases one by one.

\subsection{The Periodic Solutions When $B = m/(1-m)$ \label{sec:sec.3a}}

If we look at the 15 solutions obtained in the previous section, one common thing in all these solutions is that $\phi_{1}, \phi_{2}$ in these solutions always have $1+B\cn^2(\beta x,m)$ in the denominator. But in case $B = \frac{m}{(1-m)}$, then 
\begin{equation}\label{eq.70}
1+B\cn^2(\beta x, m) = \frac{\dn^2(\beta x, m)}{(1-m)}\,.
\end{equation} 
Further, since in all the solutions in the previous section $B = \frac{m\sn^2(\Delta,m)}{\dn^2(\Delta,m)}$, hence we find that for the special case of $B = \frac{m}{(1-m)}$ one has $\sn^2(\Delta,m) = 1$ so that $\Delta = \pm K(m)$. Now using the well known relations~\cite{as} 
\begin{equation}\label{eq.71}
\dn(y-\Delta)	= \dn(y+\Delta),		\quad
\sn(y-\Delta) 	= -\sn(y+\Delta), 	\quad
\cn(y-\Delta) 	= -\cn(y+\Delta),
\end{equation}
it is clear that in this special case, while we have exact solutions, these solutions cannot be reexpressed as a nontrivial superposition of a 
periodic kink-antikink or superposition of two periodic kinks or two periodic pulse solutions. 

Let us now analyze the 15 exact solutions in this special case. We find that in this special case, $\phi_1, \phi_2$ in the three solutions IV, VIII and X are all in terms of (inverse) Lam\'e polynomials of order one, i.e. $\frac{cn(\beta x, m)}{\dn(\beta x, m)}, \frac{\sn(\beta x, m)}{\dn(\beta x, m)}, \frac{1}{\dn(\beta x,m)}$. On the other hand, we find that $\phi_1$ and $\phi_2$ in the three solutions III, V and XII are all in terms of the (inverse) Lam\'e polynomials of order two, i.e. $\frac{cn(\beta x, m)}{\dn^2(\beta x, m)}, \frac{\sn(\beta x, m)}{\dn^2(\beta x, m)}, \frac{\cn(\beta x, m)\sn(\beta x, m)}{\dn^2(\beta x,m)}$. Remarkably, the remaining 9 solutions have an unusual form with one of $\phi_1$ and $\phi_2$ is (inverse) Lam\'e polynomial of order one as given above, while the other one is
the (inverse) Lam\'e polynomial of order two as given above. We now present exact solutions in all the fifteen cases. \\

{\bf Case I: Three Solutions in Terms of (inverse) Lam\'e Polynomials of Order One} \\ 

{\bf Solution I}

It is easy to check that
\begin{equation}\label{eq.72}
\phi_1 = \frac{A}{\dn(\beta x, m)},	\quad
\phi_2 = \frac{D\sqrt{m} \sn(\beta x, m)}{\dn(\beta x, m)},
\end{equation}
provided
\begin{subequations}
\begin{align}
\label{eq.73.1}
b_1 A^2 + d_1 D^2	&	=	b_2 A^2 + d_2 D^2 = -2(1-m) \beta^2, \\
\label{eq.73.2}
a_1 -d_1 D^2			&	= (2-m)\beta^2\,,~~a_2 - d_2 D^2 = \beta^2.
\end{align}
\end{subequations}
Note that this solution is only valid for $0 < m < 1$. In fact this is true for all the fifteen solutions presented below as well as for the six solutions presented in Appendix~\ref{sec:app.2}. 

{\bf Solution II}

It is easy to check that
\begin{equation}\label{eq.74}
\phi_1 = \frac{A}{\dn(\beta x, m)}, 	\quad
\phi_2 =	 \frac{D\sqrt{m} \cn(\beta x, m)}{\dn(\beta x, m)},
\end{equation}
is an exact solution of the coupled Eq.~\eqref{eq.1.1} and Eq.~\eqref{eq.1.2} provided 
\begin{subequations}
\begin{align}
\label{eq.75.1}
b_1 A^2 -(1-m) d_1 D^2 &	= b_2 A^2 -(1-m) d_2 D^2 = -2(1-m) \beta^2, \\
\label{eq.75.2}
a_1 +d_1 D^2 &	= (2-m)\beta^2, \quad a_2 + d_2 D^2 = (1-m) \beta^2.
\end{align}
\end{subequations}

{\bf Solution III}

It is easy to check that
\begin{equation}\label{eq.76}
\phi_1 = \frac{A\sqrt{m} \sn(\beta x, m)}{\dn(\beta x, m)}, \quad
\phi_2 = \frac{D\sqrt{m} \cn(\beta x, m)}{\dn(\beta x, m)},
\end{equation}
is an exact solution of the coupled Eq.~\eqref{eq.1.1} and Eq.~\eqref{eq.1.2} provided 
\begin{subequations}
\begin{align}
\label{eq.77.1}
b_1 A^2 -(1-m) d_1 D^2 &	= b_2 A^2 -(1-m) d_2 D^2 = -2(1-m) \beta^2, \\
\label{eq.77.2}
a_1 +m d_1 D^2 &	= (2m-1)\beta^2, \quad	a_2 +m d_2 D^2 = -(1-m) \beta^2.
\end{align}
\end{subequations}

{\bf Case II: Three Solutions in Terms of (inverse) Lam\'e Polynomials of Order Two} \\

{\bf Solution IV}

It is easy to check that
\begin{equation}\label{eq.78}
\phi_1 = \frac{A\sqrt{m} \sn(\beta x, m)}{\dn^2(\beta x, m)}, \quad
\phi_2 = \frac{D\sqrt{m} \cn(\beta x, m)}{\dn^2(\beta x, m)},
\end{equation}
is an exact solution of the coupled Eq.~\eqref{eq.1.1} and Eq.~\eqref{eq.1.2} provided 
\begin{subequations}
\begin{align}
\label{eq.79.1}
&	b_1 = b_2,	\quad
d_1 = d_2,	\quad
m b_1 A^2 = -6(1-m)^2 \beta^2,	\quad
m d_1 D^2 = -6(1-m)\beta^2, \\
\label{eq.79.2}
&	a_1 = (5-m)\beta^2,	\quad
a_2 = (5-4m)\beta^2.
\end{align}
\end{subequations}

{\bf Solution V}

It is easy to check that
\begin{equation}\label{eq.80}
\phi_1 = \frac{A\sqrt{m} \sn(\beta x, m)}{\dn^2(\beta x, m)},	\quad
\phi_2 = \frac{D\sqrt{m} \cn(\beta x, m)\sn(\beta x, m)}{\dn^2(\beta x, m)},
\end{equation}
is an exact solution of the coupled Eq.~\eqref{eq.1.1} and Eq.~\eqref{eq.1.2} provided 
\begin{subequations}
\begin{align}
\label{eq.81.1}
&	b_1 = b_2, \quad
d_1 = d_2,	\quad
b_1 A^2 = -6(1-m)^2 \beta^2,	\quad
d_1 D^2 = -6(1-m)\beta^2, \\
\label{eq.81.2}
&	a_1 = (5m-1)\beta^2,	\quad
a_2 = (5m-4)\beta^2.
\end{align}
\end{subequations}

{\bf Solution VI}

It is easy to check that
\begin{equation}\label{eq.82}
\phi_1 = \frac{A\sqrt{m} \cn(\beta x, m)}{\dn^2(\beta x, m)},	\quad
\phi_2 = \frac{D\sqrt{m} \cn(\beta x, m)\sn(\beta x, m)}{\dn^2(\beta x, m)},
\end{equation}
is an exact solution of the coupled Eq.~\eqref{eq.1.1} and Eq.~\eqref{eq.1.2} provided 
\begin{equation}\label{eq.83}
b_1 = b_2 = -d_1 = -d_2,	\quad
D = \pm A,	\quad
b_1 A^2 = 6 \beta^2,	\quad
a_1 = -(4m+1)\beta^2,	\quad
a_2 = -(4+m)\beta^2.
\end{equation}

We might add here that there is also a fourth (inverse) Lam\'e polynomial of order two, i.e. $A[\frac{1}{\dn^2(\beta x, m)}+p]$, where $p$ is a constant, which has not occured in any of the 15 solutions that we have considered. Instead, it would follow if we had considered $\phi_1 = \frac{A\cn^2(\beta x,m)}{1+B\cn^2(\beta x,m)}$. However, we did not consider it as it cannot be re-expressed as a superposition of a periodic kink-antikink or two periodic kinks or two periodic pulse solutions. For completeness, we present the three exact solutions of this (inverse) Lam\'e polynomial of order two with the other three  (inverse) Lam\'e polynomials of order two in Appendix~\ref{sec:app.2}. \\

{\bf Case III: Nine Solutions in Terms of (inverse) Lam\'e Polynomials of Order One and Two} \\

If we rearrange so that in the remaining nine solutions $\phi_1$ is always (inverse)  Lam\'e polynomial of order two while $\phi_2$ is (inverse) Lam\'e polynomial of order one, then one finds that in all the nine solutions $b_1 = b_2 = 0$ and hence effectively we need to solve rather unusual simpler coupled $\phi^4$ equations as
\begin{subequations}
\begin{align}
\label{eq.84.1}
&	\phi_{1xx} = a_1 \phi_1 + d_1 \phi_{2}^2 \phi_1, \\
\label{eq.84.2}
&	\phi_{2xx} = a_2 \phi_2 + d_2 \phi_{2}^3.
\end{align}
\end{subequations}
So far as we are aware of, such solutions have never been discussed in the literature. We now present these one by one. \\

{\bf Solution VII}

It is easy to check that
\begin{equation}\label{eq.85}
\phi_1 = \frac{A\sqrt{m}\cn(\beta x,m)}{\dn^2(\beta x,m)},	\quad	\phi_2 = \frac{D\sqrt{m}\cn(\beta x,m)}{\dn(\beta x,m)},
\end{equation}
is an exact solution of coupled Eq.~\eqref{eq.84.1} and Eq.~\eqref{eq.84.2} provided 
\begin{equation}\label{eq.86}
d_1 D^2 = 6\beta^2, 	\quad	a_1 = -(4m+1)\beta^2,	\quad	d_2 D^2 = 2\beta^2,	\quad	a_2 = -(1+m)\beta^2.
\end{equation}
This solution is valid for arbitrary nonzero values of $A$ since none of the relations in Eq. (86) depends on $A$. This is in fact true for all the 9 solutions presented here as well as for similar three solutions in Appendix~\ref{sec:app.2}.  Note that Eqs. (84a) and (84b) are nonreciprocal equations since while the solution for $\phi_1$ depends on $\phi_2$, the latter is independent of $\phi_1$. This is in fact true for all the 9 solutions presented here as well as for the similar 3 solutions presented in Appendix~\ref{sec:app.2}. \\

{\bf Solution VIII}

It is easy to check that
\begin{equation}\label{eq.87}
\phi_1 = \frac{A\sqrt{m}\cn(\beta x,m)}{\dn^2(\beta x,m)},	\quad	\phi_2 = \frac{D\sqrt{m}\sn(\beta x,m)}{\dn(\beta x,m)},	
\end{equation}
is an exact solution of coupled Eq.~\eqref{eq.84.1} and Eq.~\eqref{eq.84.2} provided 
\begin{equation}\label{eq.88}
d_1 D^2 = -6(1-m)\beta^2,	\quad	a_1 = (2m-1)\beta^2,	\quad	d_2 D^2 = -2(1-m)\beta^2,	\quad	a_2 = (2m-1)\beta^2.
\end{equation}

{\bf Solution IX}

It is easy to check that
\begin{equation}\label{eq.89}
\phi_1 = \frac{A\sqrt{m}\cn(\beta x,m)}{\dn^2(\beta x,m)},	\quad	\phi_2 = \frac{D}{\dn(\beta x,m)},
\end{equation}
is an exact solution of coupled Eq.~\eqref{eq.84.1} and Eq.~\eqref{eq.84.2} provided 
\begin{equation}\label{eq.90}
d_1 D^2 = -6(1-m)\beta^2, \quad	a_1 = (5-4m)\beta^2,	\quad	d_2 D^2 = -2(1-m)\beta^2, \quad	a_2 = (2-m)\beta^2.
\end{equation}

{\bf Solution X}

It is easy to check that
\begin{equation}\label{eq.91}
\phi_1 = \frac{A\sqrt{m}\sn(\beta x,m)}{\dn^2(\beta x,m)}, \quad \phi_2 = \frac{D\sqrt{m}\cn(\beta x,m)}{\dn(\beta x,m)}
\end{equation}
is an exact solution of coupled Eq.~\eqref{eq.84.1} and Eq.~\eqref{eq.84.2} provided 
\begin{equation}\label{eq.92}
d_1 D^2 = 6\beta^2,	\quad	a_1 = -(m+1)\beta^2,	\quad	d_2 D^2 = 2\beta^2, \quad	a_2 = -(1+m)\beta^2.
\end{equation}

{\bf Solution XI}

It is easy to check that
\begin{equation}\label{eq.93}
\phi_1 = \frac{A\sqrt{m}\sn(\beta x,m)}{\dn^2(\beta x,m)}, \quad	\phi_2 = \frac{D\sqrt{m}\sn(\beta x,m)}{\dn(\beta x,m)},
\end{equation}
is an exact solution of coupled Eq.~\eqref{eq.84.1} and Eq.~\eqref{eq.84.2} provided 
\begin{equation}\label{eq.94}
d_1 D^2 = -6(1-m)\beta^2,	\quad	a_1 = (5m-1)\beta^2,	\quad	d_2 D^2 = -2(1-m)\beta^2,	\quad	a_2 = (2m-1)\beta^2.
\end{equation}

{\bf Solution XII}

It is easy to check that
\begin{equation}\label{eq.95}
\phi_1 = \frac{A\sqrt{m}\sn(\beta x,m)}{\dn^2(\beta x,m)},		\quad	\phi_2 = \frac{D}{\dn(\beta x,m)},
\end{equation}
is an exact solution of coupled Eq.~\eqref{eq.84.1} and Eq.~\eqref{eq.84.2} provided 
\begin{equation}\label{eq.96}
d_1 D^2 = -6(1-m)\beta^2,	\quad	a_1 = (5-m)\beta^2,	\quad	d_2 D^2 = -2(1-m)\beta^2, 	\quad	a_2 = (2-m)\beta^2.
\end{equation}

{\bf Solution XIII}

It is easy to check that
\begin{equation}\label{eq.97}
\phi_1 = \frac{Am\sn(\beta x,m)\cn(\beta x,m)}{\dn^2(\beta x,m)}, 	\quad	\phi_2 = \frac{D\sqrt{m}\cn(\beta x,m)}{\dn(\beta x,m)},
\end{equation}
is an exact solution of coupled Eq.~\eqref{eq.84.1} and Eq.~\eqref{eq.84.2} provided 
\begin{equation}\label{eq.98}
d_1 D^2 = 6\beta^2,\quad	a_1 = -(4+m)\beta^2,	\quad	d_2 D^2 = 2\beta^2, 	\quad	a_2 = -(1+m)\beta^2.
\end{equation}

{\bf Solution XIV}

It is easy to check that
\begin{equation}\label{eq.99}
\phi_1 = \frac{Am\sn(\beta x,m)\cn(\beta x,m)}{\dn^2(\beta x,m)}, 	\quad \phi_2 = \frac{D\sqrt{m}\sn(\beta x,m)}{\dn(\beta x,m)},
\end{equation}
is an exact solution of coupled Eq.~\eqref{eq.84.1} and Eq.~\eqref{eq.84.2} provided 
\begin{equation}\label{eq.100}
d_1 D^2 = -6(1-m)\beta^2, \quad	a_1 = (5m-4)\beta^2,	\quad	d_2 D^2 = -2(1-m)\beta^2,	\quad	a_2 = (2m-1)\beta^2.
\end{equation}

{\bf Solution XV}

It is easy to check that
\begin{equation}\label{eq.101}
\phi_1 = \frac{Am\sn(\beta x,m)\cn(\beta x,m)}{\dn^2(\beta x,m)}, \quad	\phi_2 = \frac{D}{\dn(\beta x,m)},
\end{equation}
is an exact solution of coupled Eq.~\eqref{eq.84.1} and Eq.~\eqref{eq.84.2} provided 
\begin{equation}\label{eq.102}
d_1 D^2 = -6(1-m)\beta^2, 	\quad	a_1 = (2-m)\beta^2,	\quad	d_2 D^2 = -2(1-m)\beta^2,	\quad	a_2 = (2-m)\beta^2.
\end{equation}

For completeness we have presented the three solutions corresponding to the fourth inverse Lam\'e polynomial of order two with the three Lam\'e polynomials of order one in Appendix~\ref{sec:app.2}, even though they do not follow from any of the 15 solutions that we have considered. Instead, it would follow if we had considered $\phi_1 = \frac{A\cn^2(\beta x,m)}{1+B\cn^2(\beta x,m)}$. However, we did not consider it as it cannot be re-expressed as a superposition of a periodic kink-antikink or two periodic kinks or two periodic pulse solutions.

One obvious question (though not related with the present paper) is: are there similar solutions of coupled Eq.~\eqref{eq.84.1} and Eq.~\eqref{eq.84.2} in terms of Lam\'e polynomials of order one and two. The answer is in the affirmative. 
In fact there are 12 such periodic solutions which in turn lead to 6 hyperbolic nonreciprocal solutions. However, we will not present them in this paper.

\subsection{The Trigonometric Solutions: When $m = 0$ \label{sec:sec.3b}}

In case $m = 0$,  the addition theorems for $\sin(x)$ and $\cos(x)$ are rather simple and well known and clearly there are no superposed solutions in this limit. One way to discern it is that, in this case $\Delta = 0$. It turns out that in the limit $m = 0$, 13 out of the 15 superposed solutions discussed in the last section reduce to 6 nontrivial trigonometric solutions while solutions VII and XI are not valid in this limit. We now discuss the allowed 6 solutions one by one. \\

{\bf Solution I}

It is easy to check that
\begin{equation}\label{eq.103}
\phi_1 = \frac{A\cos(\beta x)}{1+B\cos^2(\beta x)}, \quad	\phi_2 = \frac{D\sin(\beta x)}{1+B\cos^2(\beta x)},	\quad	B > 0,
\end{equation}
is an exact solution of coupled Eq.~\eqref{eq.1.1} and Eq.~\eqref{eq.1.2} provided 
\begin{equation}\label{eq.104}
a_1 = a_2 = -\beta^2,	\quad	d_1 D^2 = 3 d_2 D^2 = -6B\beta^2,		\quad	b_2 A^2 = 3 b_1 A^2 = 6B(B+1) \beta^2.
\end{equation}
For all the solutions given below, $B > 0$ and so we avoid mentioning it again. \\

{\bf Solution II}

It is easy to check that
\begin{equation}\label{eq.105}
\phi_1 = \frac{A\cos(\beta x)}{1+B\cos^2(\beta x)}, \quad	\phi_2 = \frac{D}{1+B\cos^2(\beta x)},
\end{equation}
is an exact solution of coupled Eq.~\eqref{eq.1.1} and Eq.~\eqref{eq.1.2} provided 
\begin{subequations}
\begin{align}
\label{eq.106.1}
&	a_1 = -\beta^2,	\quad	d_1 D^2 = -6B\beta^2,	\quad	b_1 A^2 = 2B(B+4)\beta^2, \\
\label{eq.106.2}
&	a_2 = -4\beta^2,	\quad	d_2 D^2 = 2(2-B)\beta^2,	\quad	b_2 A^2 = 6B(B+2) \beta^2.
\end{align}
\end{subequations}

{\bf Solution III}

It is easy to check that
\begin{equation}\label{eq.107}
\phi_1 = \frac{A}{1+B\cos^2(\beta x)},	\quad	\phi_2 = \frac{D\sin(\beta x)}{1+B\cos^2(\beta x)},
\end{equation}
is an exact solution of coupled Eq.~\eqref{eq.1.1} and Eq.~\eqref{eq.1.2} provided 
\begin{subequations}
\begin{align}
\label{eq.108.1}
&	a_1 = -4\beta^2,	\quad	d_1 D^2 =  -6B(B+2)\beta^2,	\quad	b_1 A^2 = 2(B+1)(3B+2) \beta^2, \\
\label{eq.108.2}
&	a_2 = -\beta^2,	\quad	d_2 D^2 = -2B(3B+4)\beta^2,	\quad	b_2 A^2,
= 6B(B+1)\beta^2\,.
\end{align}
\end{subequations}

{\bf Solution IV}

It is easy to check that
\begin{equation}\label{eq.109}
\phi_1 = \frac{A}{1+B\cos^2(\beta x)},	\quad		\phi_2 = \frac{D\sin(\beta x)\cos(\beta x)}{1+B\cos^2(\beta x)},
\end{equation}
is an exact solution of coupled Eq.~\eqref{eq.1.1} and Eq.~\eqref{eq.1.2} provided 
\begin{equation}\label{eq.110}
a_1 = a_2 = 2\beta^2,	\quad	d_1 D^2 = 3d_2 D^2 =  6B^2 \beta^2,	\quad	b_2 A^2 = 3b_1 A^2 = -6(B+1) \beta^2.
\end{equation}

{\bf Solution V}

It is easy to check that
\begin{equation}\label{eq.111}
\phi_1 = \frac{A\cos(\beta x)}{1+B\cos^2(\beta x)},	\quad	\phi_2 = \frac{D\sin(\beta x)\cos(\beta x)}{1+B\cos^2(\beta x)},
\end{equation}
is an exact solution of coupled Eq.~\eqref{eq.1.1} and Eq.~\eqref{eq.1.2} provided 
\begin{subequations}
\begin{align}
\label{eq.112.1}
a_1 = -6(B+1)\beta^2,	\quad	d_1 D^2 =  -6B^3 \beta^2,	\quad	b_1 A^2 = 2B(B+1)(3B+4) \beta^2,	\\
\label{eq.112.2}
a_2 = -6(B+4)\beta^2,	\quad	d_2 D^2 =  -2(3B+2)\beta^2,	\quad	b_2 A^2 = 6B(B+1)(B+2) \beta^2.
\end{align}
\end{subequations}

{\bf Solution VI}

It is easy to check that
\begin{equation}\label{eq.113}
\phi_1 = \frac{A\sin(\beta x)}{1+B\cos^2(\beta x)},	\quad	\phi_2 = \frac{D\sin(\beta x)\cos(\beta x)}{1+B\cos^2(\beta x)},
\end{equation}
is an exact solution of coupled Eq.~\eqref{eq.1.1} and Eq.~\eqref{eq.1.2} provided 	
\begin{subequations}
\begin{align}
\label{eq.114.1}
(1+B)a_1 = (5B-1)\beta^2,	\quad	(1+B)d_1 D^2 =  6B^3 \beta^2,			\quad	(1+B) b_1 A^2 = -2B(B+4)\beta^2, \\
\label{eq.114.2}
(1+B)a_2 = 2(B-2)\beta^2,	\quad	(1+B)d_2 D^2 =  2B^2(B-2) \beta^2,	\quad	(1+B) b_2 A^2 = -6B(B+2)\beta^2.
\end{align}
\end{subequations}

Having obtained the 15 solutions of the coupled $\phi^4$ model with six parameters, $a_1, a_2, b_1, b_2, d_1, d_2$, it is worthwhile asking: is there a range of these 6 parameters for which the exact solution has not been obtained so far? We believe that these six parameters with each of them being positive and negative span a six parameter space and it is highly unlikely that one has found solutions in this entire space of parameters. Out of the 15 solutions that we have obtained, we can try to draw some conclusions by first carefully looking at the superposed solutions at $m = 1$ (i.e. hyperbolic superposed solutions) obtained earlier~\cite{ks22b} and mentioned in the previous section as well as the exact solutions when $m = (1-m)B$ and $m=0$. 

From the six hyperbolic superposed solutions mentioned in the previous section as well as the 9 solutions obtained above when $m = (1-m)B$ and the six solutions that we have obtained at $m = 0$ we note that in none of the 21 solutions either all the six parameters $a_1, a_2, b_1, b_2, d_1, d_2$ are positive or all negative. We have also looked at the 15 periodic superposed solutions obtained in the last section and for a few values of $B$ and $m$ that we have seen, we obtain the same conclusion. 

It is worthwhile looking at the other known solutions of this coupled model and see if the above conclusion is correct or not. 

\section{Novel periodic superposed solutions of a coupled NLS model \label{sec:sec.4}}

Let us consider the following coupled NLS model~\cite{ks22b} 
\begin{subequations}
\begin{align}
\label{eq.115.1}
iu_{1t}+u_{1xx} +6[g_{11}|u_1|^2+g_{12}|u_2|^2]u_{1} &	= 0, \\
\label{eq.115.2}
iu_{2t}+u_{2xx} +6[g_{21}|u_1|^2+g_{22}|u_2|^2]u_{2} &	= 0.
\end{align}
\end{subequations}
It is worth pointing out that in case $g_{11}=g_{12}=g_{21}=g_{22}$, then it corresponds to the celebrated Manakov system which is known to be integrable~\cite{man}. On the other hand, in case $g_{11} = g_{21} = -g_{12} = -g_{22}$, then it corresponds to the Manakov-Zakharov-Schulman (MZS) system~\cite{mik,zak,ger}. 

Before we discuss the superposed periodic kink and pulse solutions of Eq.~\eqref{eq.115.1} and Eq.~\eqref{eq.115.2}, let us note that these coupled equations also admit periodic kink and pulse solutions. 

We now show that these coupled equations admit 19 periodic solutions which can be rewritten as superposed solutions either of the form $\cn(\beta x+\Delta) \pm \cn(\beta x -\Delta)$\,,~ or $\dn(\beta x+\Delta) \pm \dn(\beta x -\Delta)$\,, ~or
$\sn(\beta x+\Delta) \pm \sn(\beta x -\Delta)$\,. 

We start from Eqs. (\ref{eq.115.1}) and (\ref{eq.115.2}) and make an ansatz
\begin{equation}\label{eq.116}
u_{1}(x,t) = e^{-i \omega_1 t} u_1(x), \quad	u_2(x,t) = e^{-i \omega_2 t} u_2(x),
\end{equation}
so that the two coupled equations take the form
\begin{subequations}
\begin{align}
\label{eq.117.1}
&	u_{1xx}(x) = \omega_1 u_1 -6[g_{11}|u_1|^2 +g_{12}|u_2|^2]u_1, \\
\label{eq.117.2}
&	u_{2xx}(x) = \omega_2 u_2 -6[g_{21}|u_1|^2 +g_{22}|u_2|^2]u_2.
\end{align}
\end{subequations}
Observe that in case $u_1(x)$ and $u_2(x)$ are real, the coupled equations [Eq.~\eqref{eq.117.1}, and Eq.~\eqref{eq.117.2}] can be directly mapped to the coupled $\phi^4$ equations [Eq.~\eqref{eq.1.1} and Eq.~\eqref{eq.1.2}] with the identification of $\omega_1, \omega_2$ with $a_1, a_2$, respectively. Further, the four coupling constants $g_{11}, g_{12}, g_{21}, g_{22}$ can be identified with $-b_1, -d_1, -b_2, -d_2$, respectively. One can then simply read off the 19 superposed periodic solutions of the coupled NLS model from the corresponding 15 superposed periodic solutions of the coupled $\phi^4$ model as obtained in Sec. II and 4 superposed solutions of the coupled $\phi^4$ model as obtained in Appendix~\ref{sec:app.1}. For completeness, we simply mention one solution of the coupled NLS model and the readers can similarly obtain the remaining 18 superposed solutions of the coupled NLS model including the 4 given in Appendix~\ref{sec:app.1}.

\subsection{Solutions of coupled NLS when $u_1(x,t)$ and $u_2(x,t)$ are distinct \label{sec:sec.5}}

{\bf Solution I}

It is easy to check that
\begin{equation}\label{eq.118}
u_1(x,t) = e^{i\omega_1 t} \frac{A\cn(\beta x,m)}{1+B\cn^2(\beta x,m)}, \quad
u_2(x,t) = e^{i\omega_2 t} \frac{D \sn(\beta x,m)\dn(\beta x,m)} {1+B\cn^2(\beta x,m)},
\end{equation}
with $B >0$ is an exact solution of the coupled equations, Eq.~\eqref{eq.115.1} and Eq.~\eqref{eq.115.2}, provided
\begin{align}
\nonumber
&	\omega_1 = \omega_2 = (2m-1)\beta^2, \quad	g_{12} D^2 = 3 g_{22} D^2 = 6 B\beta^2, \\
\label{eq.119}
&	g_{21} A^2  = 3 g_{11} A^2 = 6(B+1)[m-(1-m)B]\beta^2.
\end{align}
Notice that for this solution $g_{12}, g_{22} >$ 0. Further, $\omega_1, \omega_2 \ge (<)$ 0 in case $m \ge (<)$ 1/2 while $g_{12}, g_{22} > 0$. Finally, $g_{11}, g_{21} \ge (<)$ 0 depending on if $m \ge (<)$ $(1-m)B$.

On using the identities Eq.~\eqref{eq.5.1} and Eq.~\eqref{eq.5.2}, the coupled solution Eq.~\eqref{eq.118} can be rewritten as 
\begin{subequations}
\begin{align}
\label{eq.120.1}
&	u_1(x,t) = e^{i\omega_1 t} \sqrt{\frac{m}{2g_{11}}}\beta  [\cn(\beta x+\Delta,m)+\cn(\beta x-\Delta,m)], \\
\label{eq.120.2}
&	u_2(x,t) = e^{i\omega_2 t} \sqrt{\frac{m}{2g_{22}}} \beta [\cn(\beta x-\Delta,m)  - \cn(\beta x+\Delta,m)],
\end{align}
\end{subequations}
where $B = \frac{m\sn^2(\Delta,m)}{\dn^2(\Delta,m)}$. 

\section{Conclusion and Open Problems \label{sec:sec.6}}

In this article, we have expanded the concept of superposition to encompass coupled $\phi^4$ and coupled NLS models, unveiling their capacity to exhibit not only hyperbolic solutions but also periodic solutions. These periodic solutions can be expressed as a superposition of a periodic kink and an antikink, two periodic kinks, or two periodic pulse solutions. Remarkably, we have identified fifteen such superposed periodic solutions when the coupled fields are distinct (and not proportional to each other), and four periodic solutions when the fields are proportional.

Our findings raise intriguing questions that warrant further exploration. We highlight a few of these questions:

\begin{enumerate}

\item Unlike the coupled $\phi^4$ or coupled NLS models, we have been unable to obtain periodic solutions for the coupled mKdV equation that can be expressed as a superposition of a periodic kink and an antikink, or two periodic kinks, or two periodic pulse solutions. While coupled hyperbolic solutions of the mKdV equation have been obtained in a different form as the superposition of a kink and an antikink~\cite{ks22a}, it is worthwhile to investigate if the concept of superposition can be extended to periodic solutions of the coupled mKdV equation as well.

\item Numerous other coupled equations, such as coupled KdV, coupled asymmetric $\phi^4$, and coupled $\phi^6$, merit exploration to determine the extent to which the notion of superposition can be applied to their periodic solutions. 

\item Is it possible to extend the concept of superposition in alternative and unexplored directions for nonlinear equations? Given the vast richness of nonlinear theories, this pursuit holds promising potential. 

\item What are the direct implications of superposed periodic kink-antikink or two periodic kink/pulse solutions? Do they correspond to bound states or merely excite the system?


\item Can the notion of superposition be extended to nonlocal theories? Our recent endeavors~\cite{ks22c,ks22d} have partially extended this concept to nonlocal NLS~\cite{abm, yang}, nonlocal mKdV~\cite{he18}, nonlocal Hirota equations~\cite{ccf,xyx} , as well as coupled nonlocal Ablowitz-Musslimani variant of NLS and coupled nonlocal mKdV models~\cite{ks14}. It is worth exploring the possibility of extending this concept to other nonlocal models.

\end{enumerate}

We intend to address some of the issues raised above in the near future.

\section{Acknowledgment \label{sec:sec.7}}

One of us (AK) is grateful to Indian National Science Academy (INSA) for the award of INSA Honorary Scientist position at Savitribai Phule Pune University. The work at Los Alamos National Laboratory was carried out under the auspices of the US DOE and NNSA under contract No.~DEAC52-06NA25396.

\appendix

\section{Superposed periodic solutions of the coupled $\phi^4$ model: $\phi_2(x) \propto \phi_1(x)$  \label{sec:app.1}}

In case $\phi_2(x) = \alpha \phi_1(x)$, with $\alpha$ being a real number, it is straightforward to verify that Eq.~~\eqref{eq.1.1} and Eq.~\eqref{eq.1.2} are consistent with each other provided
\begin{equation}\label{aeq.1}
a_2 = a_1\,,~~b_2 +\alpha^2 d_2 = b_1+\alpha^2 d_1.
\end{equation}
Thus, the relevant equation to solve in this case is
\begin{equation}\label{aeq.2}
\phi_{1xx} = a_1 \phi_1 + (b_1+\alpha^2 d_1) \phi^3.
\end{equation}
Now, in a recent publication~\cite{ks22a} we have already solved such an equation and shown that it admits four superposed solutions which we can immediately read off from that paper. For completeness we provide these solutions below. \\

{\bf Solution I}

It is readily checked that the $\phi^4$ field Eq.~\eqref{aeq.2} and hence the coupled equations Eq.~\eqref{eq.1.1} and Eq.~\eqref{eq.1.2} admit the periodic solution
\begin{equation}\label{aeq.3}
\phi_1(x) = \frac{A \dn(\beta x,m) \cn(\beta x,m)}{1+B\cn^2(\beta x,m)}, \quad
A,B,D > 0,
\end{equation}
provided Eq.~\eqref{aeq.1} is satisfied and further
\begin{align}
\nonumber
&	0 < m < 1\,,~~B = \frac{\sqrt{m}}{1-\sqrt{m}}, \\
\label{aeq.4}
&	a_1 = -[1+m+6\sqrt{m}]\beta^2 < 0, \quad (b_1+\alpha^2 d_1) A^2 = \frac{8 \sqrt{m} \beta^2}{(1-\sqrt{m})^2}.
\end{align}
Notice that for this solution $a_1 < 0, b_1+\alpha^2 d_1 > 0$. Furthermore, this solution is only valid if $0 < m < 1$, i.e. there is no corresponding superposed hyperbolic solution. On using the identity Eq.~\eqref{eq.3.2}, one can express differently the periodic solution I given by Eq.~\eqref{aeq.3} as superposition of a periodic kink and and an antikink solutions, i.e.
\begin{equation}\label{aeq.5}
\phi_1(x) = \frac{\sqrt{2m} \beta}{\sqrt{b_1+\alpha^2 d_1}} \bigg [\sn(\beta x +\Delta, m) -\sn(\beta x- \Delta, m) \bigg ].
\end{equation}
Here $\Delta$ is defined by $\sn(\sqrt{m}\Delta,1/m) = \pm m^{1/4}$, where the following identity has been used
\begin{equation}\label{aeq.6}
\sqrt{m} \sn(y,m) = \sn(\sqrt{m} y,1/m).
\end{equation}

{\bf Solution II}

Remarkably, the $\phi^4$ Eq.~\eqref{aeq.2} also admits another periodic solution
\begin{equation}\label{aeq.7}
\phi_1(x) = \frac{A \sn(\beta x,m)}{1+B\cn^2(\beta x,m)}\,,~~A,B,D > 0,
\end{equation}
provided
\begin{equation}\label{eqaeq.8}
0 < m < 1,	\quad	B = \frac{\sqrt{m}}{1-\sqrt{m}}, 	\quad	a_1 = [6\sqrt{m}-(1+m)]\beta^2,	\quad	(b_1+\alpha^2 d_1) A^2 = -8\sqrt{m} \beta^2.
\end{equation}
Thus for this solution while $b_1+\alpha^2 d_1 < 0$, $a_1 > (<)$ 0 depending on
if $6\sqrt{m} > (<)$ 0. On using the identity Eq.~\eqref{eq.3.1}, the solution Eq.~\eqref{aeq.7} can be expressed differently as a superposition of two periodic kink solutions, i.e.
\begin{equation}\label{aeq.9}
\phi_1(x) = i\sqrt{\frac{2m}{|b_1+\alpha^2 d_1|}}\beta \bigg [\sn(\beta x +\Delta, m)+\sn(\beta x -\Delta, m) \bigg ].
\end{equation}
Here $\Delta$ is defined by $\sn(\sqrt{m}\Delta,1/m) = \pm m^{1/4}$, where we utilize the identity Eq.~\eqref{aeq.6}.

It is worth to note that for both the solutions I and II, the value of $B$ is the same but while $b_1+\alpha^2 d_1 > 0$ for the first solution, $b_1+\alpha^2 d_1 <0$ for the second solution. Further, both the solutions are valid if $0 < m < 1$. \\

{\bf Solution III}

It is readily checked that the $\phi^4$ field Eq.~\eqref{aeq.2} admits another periodic solution
\begin{equation}\label{aeq.10}
\phi_1(x) = \frac{A \sn(\beta x,m) \cn(\beta x,m)}{1+B\cn^2(\beta x,m)}, \quad
A,B,D > 0,
\end{equation}
provided
\begin{equation}\label{aeq.11}
0 < m < 1,	\quad
B = \frac{1-\sqrt{1-m}}{\sqrt{1-m}},	\quad
a_1 = (2-m-6\sqrt{1-m})\beta^2,	\quad
(b_1+\alpha^2 d_1) A^2 = -\frac{8(1-\sqrt{1-m})^2 \beta^2}{\sqrt{1-m}}.
\end{equation}
Note that for this solution $b_1+\alpha^2 d_1 < 0$ while $a_1$ could be positive or negative depending on the value of $m$. On using the identity Eq.~\eqref{eq.6.2}, the solution Eq.~\eqref{aeq.10} can be expressed differently as a superposition of two periodic $\dn(x,m)$ solutions, i.e.
\begin{equation}\label{aeq.12}
\phi_1(x) = \beta \sqrt{\frac{2}{|b_1+\alpha^2 d_1|}} \bigg (\dn[\beta x -\frac{K(m)}{2},m] - \dn[\beta x +\frac{K(m)}{2},m] \bigg ).
\end{equation}

{\bf Solution IV}

Remarkably, the $\phi^4$ Eq.~\eqref{aeq.2} also allows another periodic solution
\begin{equation}\label{aeq.13}
\phi_1(x) = \frac{A \dn(\beta x,m)}{1+B\cn^2(\beta x,m)}, \quad 
A,B,D > 0,
\end{equation}
provided
\begin{equation}\label{aeq.14}
0 < m < 1,	\quad
B =  \frac{1-\sqrt{1-m}}{\sqrt{1-m}},	\quad
a_1 = [2-m+6\sqrt{1-m}]\beta^2,	\quad
(b_1+\alpha^2 d_1) A^2 = -\frac{8}{\sqrt{1-m}} \beta^2.
\end{equation}
Thus for this solution while $b_1+\alpha^2 d_1 < 0$, $a_1 > 0$. On employing the identity Eq.~\eqref{eq.6.1}, the periodic solution Eq.~\eqref{aeq.13} can be expressed differently as a superposition of two periodic $\dn(x,m)$ pulse solutions, i.e.
\begin{equation}\label{aeq.15}
\phi_1(x) = \sqrt{\frac{2}{|b_1+\alpha^2 d_1|}} \beta  \bigg (\dn[\beta x +K(m)/2, m]+\dn[\beta x -K(m)/2, m] \bigg ) \,, 
\end{equation}
where $K(m)$ is the complete elliptic integral of the first kind. 

It is worth to note that for both the superposed periodic pulse solutions III and IV, not only the value of $B$ is the same but also $b_1+\alpha^2 d_1 < 0$ for both the solutions. Further, both of them are not valid for $m = 1$, i.e. there is no hyperbolic superposed pulse solution.

It is also worth pointing out that while the solutions I to IV admit superposed periodic $\sn(x,m)$ kink solutions and superposed periodic $\dn(x,m)$ pulse solutions, Eq.~\eqref{aeq.2} does not admit superposed $\cn(x,m)$ solutions. However, as shown in Sec.~\ref{sec:sec.2}, in case the two fields $\phi_1$ and $\phi_2$ are distinct then superposed solutions of $\cn(x,m)$ type are also admitted.

\section{Some Exact Solutions of the (Inverse) Lam\'e Polynomials of Order II  \label{sec:app.2}}

We first present three exact solutions of the fourth (inverse) Lam\'e 
polynomial, i.e. $A[\frac{1}{\dn^2(\beta x, m)}+p]$ with the other three 
(inverse) Lam\'e polynomials of order two. \\

{\bf Solution I}

It is easy to check that
\begin{equation}\label{beq.1}
\phi_1 = A\left[\frac{1}{\dn^2(\beta x, m)}+p\right], \quad	\phi_2 = \frac{D\sqrt{m}\sn(\beta x, m)}{\dn^2(\beta x, m)},
\end{equation}
is an exact solution of the coupled equations [Eq.~\eqref{eq.1.1} and Eq.~\eqref{eq.1.1}] provided
\begin{subequations}
\begin{align}
\label{beq.2.1}
&	b_1 = b_2,	\quad	d_1 = d_2,	\quad	b_1 A^2 = -d_1 D^2,	\quad	(2p+1)b_1 A^2 =-6(1-m)\beta^2, \\
\label{beq.2.2}
&	a_1 = \beta^2[-\frac{2}{p}-p^2 b_1 A^2], 	\quad	a_2 = \beta^2[(5-m)-p^2 b_1 A^2],
\end{align}
\end{subequations}
where
\begin{equation}\label{beq.3}
p = \frac{-(2-m)\pm \sqrt{1-m+m^2}}{3(1-m)}.
\end{equation}

{\bf Solution II}

It is easy to check that
\begin{equation}\label{beq.4}
\phi_1 = A\left[\frac{1}{\dn^2(\beta x, m)}+p\right],	\quad	\phi_2 = \frac{D\sqrt{m}\cn(\beta x, m)}{\dn^2(\beta x, m)},
\end{equation}
is an exact solution of the coupled equations [Eq.~\eqref{eq.1.1} and Eq.~\eqref{eq.1.1}] provided
\begin{subequations}
\begin{align}
\label{beq.5.1}
&	b_1 = b_2,	\quad	d_1 = d_2,	\quad	b_1 A^2 = (1-m)d_1 D^2,	\quad	[2p(1-m)+1]d_1 D^2 =-6(1-m)\beta^2, \\
\label{beq.5.2}
&	a_1 = \beta^2[-\frac{2}{p}+(1-m)p^2 d_1 D^2], \quad	a_2 = \beta^2[(5-4m)-(1-m) p^2 d_1 D^2],
\end{align}
\end{subequations}
where $p$ is again given by Eq.~\eqref{beq.3}. \\

{\bf Solution III}

It is easy to check that
\begin{equation}\label{beq.6}
\phi_1 = A\left[\frac{1}{\dn^2(\beta x, m)}+p\right],	\quad	\phi_2 = \frac{Dm\cn(\beta x, m)\sn(\beta x, m)}{\dn^2(\beta x, m)},
\end{equation}
is an exact solution of the coupled  equations [Eq.~\eqref{eq.1.1} and Eq.~\eqref{eq.1.1}] provided
\begin{subequations}
\begin{align}
\label{beq.7.1}
&	b_1 = b_2,	\quad	d_1 = d_2,	\quad	b_1 A^2 = (1-m)d_1 D^2,	\quad	[2p(1-m)+2-m]d_1 D^2 =-6(1-m)\beta^2, \\
\label{beq.7.2}
&	a_1 = \beta^2 \bigg(-\frac{2}{p}+[(1-m)p^2-1] d_1 D^2 \bigg), \quad	a_2 = \beta^2\bigg( (2-m)+[1-(1-m) p^2] d_1 D^2 \bigg),
\end{align}
\end{subequations}
where $p$ is again given by Eq.~\eqref{beq.3}. \\

We now present three exact nonreciprocal solutions of the fourth (inverse) 
Lam\'e polynomial, i.e. $A[\frac{1}{\dn^2(\beta x, m)}+p]$ with the three 
(inverse) Lam\'e polynomials of order one. \\

{\bf Solution IV}

It is easy to check that
\begin{equation}\label{beq.8}
\phi_1 = A\left[\frac{1}{\dn^2(\beta x,m)}+p\right],	\quad	\phi_2 = \frac{D\sqrt{m}\cn(\beta x,m)}{\dn(\beta x,m)},
\end{equation}
where $p$ is a number, is an exact solution of coupled  equations [Eq.~\eqref{eq.1.1} and Eq.~\eqref{eq.1.1}] provided
\begin{equation}\label{beq.9}
d_1 D^2 = 6\beta^2,	\quad	a_1 = -2(3+\frac{1}{p})\beta^2,
\end{equation}
where $p$ is again given by Eq.~\eqref{beq.3} while 
\begin{equation}\label{beq.10}
d_2 D^2 = 2\beta^2,	\quad	a_2 = -(1+m)\beta^.
\end{equation}

{\bf Solution V}

It is easy to check that
\begin{equation}\label{beq.11}
\phi_1 = A\left[\frac{1}{\dn^2(\beta x,m)}+p\right],	\quad	\phi_2 = \frac{D\sqrt{m}\sn(\beta x,m)}{\dn(\beta x,m)},
\end{equation}
is an exact solution of coupled equations [Eq.~\eqref{eq.84.1} and Eq.~\eqref{eq.84.1}] provided
\begin{equation}\label{beq.12}
d_1 D^2 = -6(1-m)\beta^2,	\quad	a_1 = -2[3(1-m)+\frac{1}{p}]\beta^2,
\end{equation}
where $p$ is again given by Eq.~\eqref{beq.3}. Further,
\begin{equation}\label{beq.13}
d_2 D^2 = -2(1-m)\beta^2,	\quad	a_2 = (2m-1)\beta^2.
\end{equation}

{\bf Solution VI}

It is easy to check that
\begin{equation}\label{beq.14}
\phi_1 = A\left[\frac{1}{\dn^2(\beta x,m)}+y\right],	\quad	\phi_2 = \frac{D}{\dn(\beta x,m)},
\end{equation}
is an exact solution of coupled equations [Eq.~\eqref{eq.84.1} and Eq.~\eqref{eq.84.1}] provided
\begin{equation}\label{beq.15}
d_1 D^2 = -6(1-m)\beta^2,	\quad	a_1 = -\frac{2}{p}\beta^2,
\end{equation}
where $p$ is again given by Eq.~\eqref{beq.3}. Further, 
\begin{equation}\label{beq.16}
d_2 D^2 = -2(1-m)\beta^2,	\quad	a_2 = (2-m)\beta^2.
\end{equation}



\end{document}